# Convective flux analysis on the instability of one-dimensional detonation

Yunfeng Liu[1,2]
1. Institute of Mechanics, Chinese Academy of Sciences, Beijing 100190, China
2. School of Engineering Science, University of Chinese Academy of Sciences, Beijing 100049, China
liuyunfeng@imech.ac.cn

**Abstract:** One-dimensional numerical simulations using the Euler equations and irreversible one-step Arrhenius kinetics are conducted to study the instability mechanism of a one-dimensional gaseous detonation. By increasing the activation energy, this study identifies the characteristics of stable detonation, periodic detonation, pulsating detonation, and detonation quenching. The key difference between this study and previous research is that it is the first quantitative analysis of convective flux, kinetic energy flux, and chemical reaction heat flux. These three fluxes undergo intensive change on the detonation front and the flow field at each time step depends on the algebra summation of them. The mechanisms of detonation instability, detonation reignition and detonation quenching process can be revealed quantitatively by analyzing these fluxes.

## 1. Introduction

Detonation is a supersonic combustion wave in which the reactants are ignited by the shock wave and the energy released from the chemical reaction is coupled with the shock wave to sustain its propagation [1]. Two important theories of steady detonation, the Chapman-Jouguet (C-J) theory and the Zeldovich-von-Neumann-Doring (ZND) theory, have been frequently used in the detonation study. However, detonation is inherently unstable and understanding detonation instability remains a key challenge in detonation theory.

One method to investigate the detonation instability is the linear stability analysis, which imposes small perturbations on the steady solution to determine whether the amplitude of the perturbations grows or not. The assumption of small perturbations allows linearization and integration of the equations, enabling the identification of unstable modes. The linear stability analysis on detonation stability was first developed by Erpenbeck to investigate the stability behavior of an idealized detonation [2-4]. With the Laplace transform approach discussed in his work, the overall linear stability of a detonation was first demonstrated mathematically by analyzing the governing equations with perturbation involved in the problem.

Lee and Stewart developed a normal mode approach to address the same problem by using a numerical shooting algorithm under the acoustic boundary condition at the end of the reaction [5]. Sharpe derived a similar approach based on the normal mode analysis and obtained an asymptotic solution of the ordinary differential equations for the linear perturbations [6]. Short conducted similar studies [7,8]. A detailed review of the normal mode stability formulation was given by Gorchkov et al. [9]. Uy et al. conducted linear stability analysis of one-dimensional detonation coupled with vibrational relaxation and the Landau-Teller model was applied to specify the vibrational relaxation [10]. Xu et al. studied the pulsating behavior and stability parameters of one-dimensional non-ideal detonations [11]. Non-ideal detonations were modeled with two-step consecutive irreversible reactions to represent endothermic effects.

As with most hydrodynamic stability analyses, the dispersion relation is complex and cannot be expressed analytically, making it difficult to elucidate the physical basis of the stability mechanism. It should be noted that linear stability analyses are valid only for the initial growth of the perturbations and cannot describe the results far from the stability limits or the final nonlinear unstable structure of the detonation [1]. Thus, it is rather difficult to obtain physical insight into the stability mechanisms. Furthermore, linear stability analyses do not reveal the gas-dynamic mechanisms behind the generation of the instability.

Another method is direct numerical simulation, which employs time-dependent, nonlinear, compressible, and reactive Euler equations. Unlike the linear stability analysis, the direct numerical simulations have the advantage that the full nonlinearity of the problem is retained. A series of numerical studies were conducted by Sharp et al. using a two-step chain-branching reaction model [12-14], Short et al. using a three-step chain-branching reaction model [15,16] and Hwang et al. [17]. Ng et al. conducted numerical simulations to study the nonlinear dynamics and chaos analysis of one-dimensional pulsating detonations [18,19]. The Euler equations with overall one-step Arrhenius kinetics were used. One notable characteristic is that varying the activation energy control parameter can cause the pulsation pattern to transition from periodic to highly irregular, or chaotic, structures.

Kasimov and Stewart proposed an improved numerical simulation method that integrates the governing equations in a shock-attached frame with a nonreflecting boundary condition [20]. Henrick et al. numerically simulated the pulsating one-dimensional detonations with fifth-order accuracy [21]. A novel, highly accurate numerical scheme based on shock-fitting coupled with fifth-order WENO scheme was applied to the classical unsteady detonation problem to generate solutions with unprecedented accuracy. As the activation energy is increased, a series of period-doubling events are predicted and the system undergoes a transition to chaos. The bifurcation points are seen to converge at a rate of the Feigenbaum constant, which is consistent with the theory of non-linear dynamics. Powers gave the detailed review of multiscale modeling of detonation [22].

Choi et al. [23] systematically discussed the effect of several numerical issues, including the effect of grid size, time step, computational domain, and boundary conditions, on the simulation of detonation cellular structure with variable stability regimes ranging from weakly to highly unstable detonations. In higher dimensions, the effect of chaos may play a prominent role in detonation instability [24]. The two-dimensional detonation cellular structure is fundamentally shaped by pulsating instability. The detonation in free space was numerically simulated to understand the evolution of the one-dimensional pulsating instability and two-dimensional cellular structure [25,26]. It is found that the pulsation occurs in three stages: (1) rapid decay of the overdrive, (2) transition to the Chapman-Jouguet state accompanied by weak pulsations, and (3) development of strong pulsations. A chemical explosive mode analysis further confirmed the highly autoignition nature of the mixture in the induction zone between reaction front and shock front where thermal diffusion plays a negligible role [27].

Although the bifurcation diagram was constructed from previous computational results, the theoretical explanation for how the instability pattern transitions to chaos remains unclear. The mechanisms of detonation quenching and the thermodynamics of deflagration wave were discussed by analyzing the convective flux and heat release flux in one-dimensional numerical simulation [28]. It was first revealed that the convective flux plays an important role in detonation instability. This study is an in-depth extension of the previous work, which discusses the mechanism of detonation instability systematically.

## 2. Governing equations and numerical methods

The governing equations are the one-dimensional Euler equations incorporating overall one-step irreversible Arrhenius chemical reaction kinetics. Viscosity, diffusion, and heat conduction are neglected. The conservation form of the mass, momentum, energy, and species equations is presented here.

$$\frac{\partial U}{\partial t}+\frac{\partial F}{\partial x}=S \tag{1}$$

where, the conservation variable $U$, the convective flux $F$ and the reactive source term $S$ are, respectively,

$$U=\begin{pmatrix}\rho\\ \rho u\\ \rho e\\ \rho Z\end{pmatrix}\quad F=\begin{pmatrix}\rho u\\ \rho u^2+p\\ (\rho e+p)u\\ \rho uZ\end{pmatrix}\quad S=\begin{pmatrix}0\\ 0\\ 0\\ \dot{\omega}\end{pmatrix} \tag{2}$$

$$e=\frac{p}{(\gamma-1)\rho}+\frac{1}{2}u^2+Zq \tag{3}$$

$$p=\rho RT \tag{4}$$

$$\dot{\omega} = -K\rho Z \exp\left(-\frac{CE_{\mathrm{a}}(\gamma-1)}{RT}\right) \tag{5}$$

$$\gamma(Z) = \frac{\gamma_1 R_1 Z/(\gamma_1-1) + \gamma_2 R_2 (1-Z)/(\gamma_2-1)}{R_1 Z/(\gamma_1-1) + R_2(1-Z)/(\gamma_2-1)} \tag{6}$$

$$R(Z) = R_1 Z + R_2(1-Z) \tag{7}$$

In the above equations, $\rho$, $p$, $u$, $e$, $q$, $R$, $\gamma$ and $Z$ are density, pressure, velocity, specific total energy, specific heat release per unit mass of reactant, gas constant, specific heat ratio and the chemical reaction progress parameter, respectively. $\dot{\omega}$ is the mass production rate of combustion products, $K$ is the pre-exponential factor, $T$ is the temperature in Kelvin *K*, and $E_a$ is the activation energy, respectively. The parameter $C$ in Eq. (5) is a controlling parameter used to change the activation energy in different numerical cases. The subscripts 1 and 2 in Eqs. (6) and (7) represent the reactants and products, respectively.

According to Eqs. (8)-(11), the pressure gain Δp at one time step is determined by three terms: the convective flux, the kinetic energy term and the heat release term. To facilitate numerical analysis, we define the convective flux ($\Gamma$), the enthalpy flux ($F_E$), the kinetic energy term ($\Delta K$), and the chemical reaction heat release flux ($\Delta Q$) at each time step. In the numerical simulations, at each time step, Γ>0 on the detonation front means shock wave and Γ<0 on the detonation front means rarefaction wave.

$$U^{n+1} = U^n - \frac{\partial F^n}{\partial x}\Delta t + S^n \Delta t \tag{8}$$

$$\Delta(\rho e) = (\rho e)^{n+1} - (\rho e)^n = -\frac{\partial(\rho e + p)u}{\partial x}\Delta t \tag{9}$$

$$\rho e = \frac{p}{(\gamma-1)} + \frac{1}{2}\rho u^2 + \rho Z q \tag{10}$$

$$\Delta p = p^{n+1} - p^n = \left(\Delta(\rho e) - \Delta\left(\frac{1}{2}\rho u^2\right) - \Delta(\rho Z q)\right)(\gamma-1) \tag{11}$$

$$F_E = (\rho e + p)u \tag{12}$$

$$\Gamma = \Delta(\rho e) = -\frac{\partial(\rho e + p)u}{\partial x}\Delta t \tag{13}$$

$$\Delta Q = -\Delta(\rho q Z) = \rho^n Z^n q - \rho^{n+1} Z^{n+1} q \tag{14}$$

$$\Delta K = -\Delta\left(\frac{1}{2}\rho u^2\right) = \frac{1}{2}\rho^n \left(u^n\right)^2 - \frac{1}{2}\rho^{n+1}\left(u^{n+1}\right)^2 \tag{15}$$

$$\Delta p = (\Gamma + \Delta K + \Delta Q)(\gamma - 1) \tag{16}$$

This methodology can be easily extended to multidimensional detonation. For instance, the convective flux and kinetic energy term of a three-dimensional laminar detonation with Navier-Stokes equations become,

$$\Gamma = \Delta(\rho e) = -\frac{\partial(\rho e + p)u_i}{\partial x_i}\Delta t + \frac{\partial}{\partial x_i}\left[\alpha \frac{\partial T}{\partial x_i} + u_j \tau_{ij}\right]\Delta t \tag{17}$$

$$\tau_{ij} = \mu\left(\frac{\partial u_i}{\partial x_j} + \frac{\partial u_j}{\partial x_i} - \frac{2}{3}\delta_{ij}\frac{\partial u_k}{\partial x_k}\right) \tag{18}$$

$$\Delta K = -\Delta\left(\frac{1}{2}\rho u_i^2\right) \tag{19}$$

In the above equations, $u_i$ is the velocity components, $\alpha$ is the laminar thermal conductivity, $\tau_{ij}$ is the viscous stress tensor, $\mu$ is the dynamic viscosity, $\delta_{ij}$ is the Kronecker symbol, respectively.

According to Eq.(9), to keep detonation propagation, the gas on the detonation front must maintain a higher enthalpy value. This means that the gas on the detonation front should obtain higher pressure gain $\Delta p$ through chemical heat release at each time step. The value of kinetic energy term ΔK in Eq.(16) is negative. Therefore, according to Eq. (16), the positive summit of convective flux Γ and the summit of heat release ΔQ should be at the same grid point at every time step in order to obtain higher enthalpy. In other words, the flame front should be coupled with the shock wave tightly. However, the position of heat release is controlled by the Arrhenius law. If the flame is decoupled from the shock wave front under higher activation energy, the peaks of heat release and shock wave are not at the same grid point, and the pressure gain by heat release becomes smaller. The physics of these fluxes at one grid point and one time step are summarized briefly in Table 1.

**Table 1. The Physics of three fluxes**

| No. | Modes | Physics of flow field |
|---|---|---|
| 1 | $\Gamma>0$, $\Delta Q>0$, $\Delta K<0$ | Total energy per unit volume is increased; heat is released; both are converted into internal energy and kinetic energy; C-J detonation. |
| 2 | $\Gamma>0$, $\Delta Q=0$, $\Delta K<0$ | Total energy per unit volume is increased; no heat release; a part of total energy is converted into kinetic energy; Shock wave. |
| 3 | $\Gamma<0$, $\Gamma+\Delta Q<0$, $\Delta K<0$ | Total energy per unit volume is reduced; the summation of convective flux and heat release is negative; Rarefaction wave. |

One-dimensional numerical simulations are conducted in this study. The computational domain is a straight detonation tube with the left-end closed and the right-end open. The detonation tube is fully filled with the premixed stoichiometric $H_2$-air mixture at 1atm and 300 K. The detonation is initiated by a smaller region of reactants with high pressure and high temperature near the closed-end wall and propagates from left to right in different numerical cases with different activation energy.

The parameters of this overall irreversible one-step Arrhenius detonation model for stoichiometric $H_2$-air mixture at initial conditions of 1atm and 300K used in this study are $Z_1 = 1.0$, $Z_2 = 0$, $\gamma_1 = 1.4$, $\gamma_2 = 1.244$, $R_1 = 398.5\mathrm{J/(kg \cdot K)}$, $R_2 = 348.9\mathrm{J/(kg \cdot K)}$, $q = 3.5\mathrm{MJ/kg}$, $E_\mathrm{a} = 4.794/(\gamma_2 - 1)\mathrm{MJ/kg}$, $K = 7.5\times10^{9}\mathrm{s^{-1}}$, respectively. This overall one-step detonation model can be found in the Ref. [29-31].

The second-order ENO scheme and third-order TVD Runge-Kutta method are implemented in the homemade code. The ENO scheme is used here in order to keep consistent with the previous study. The convective flux is split by Steger-Warming flux splitting method. The mirror reflection boundary condition is applied on the closed-end wall. The uniform grid size is 5μm and the CFL number is 0.8.

## 3. Results and discussion

### 3.1 the general characteristics of detonation

The activation energy is changed by changing the value of the controlling parameter C in different numerical cases and its value is kept constant in one case. All the results discussed in this paper are in the laboratory coordinate system. The detonation velocity (D) is defined as the propagation velocity of detonation front or the leading shock wave front and the maximum pressure ($P_{max}$) is defined as the maximum pressure of the whole flow field.

The detonation velocity and the maximum pressure ratio are plotted in Fig.1. Figure 1(a) shows the detonation propagation velocity with different values from $C$=0.6 to 1.15. If the activation energy is lower than C=0.5, the premature combustion of reactants will occur. The detonation propagation mode can be categorized into three regions roughly based on velocity: a stable region (C=0.6–0.9), an unstable region (C=1.0–1.1), and a quenching region (C⩾1.15). In the stable region, the average detonation velocity matches the theoretical C-J detonation velocity (1950 m/s) for a stoichiometric $H_2$-air mixture. The detonation velocity in the unstable region oscillates very violently between 1400m/s and 2850m/s, or between $0.71D_{C\text{-}J}$ and $1.46D_{C\text{-}J}$. The detonation is quenched abruptly to a deflagration wave when the value of parameter C is larger than 1.15.

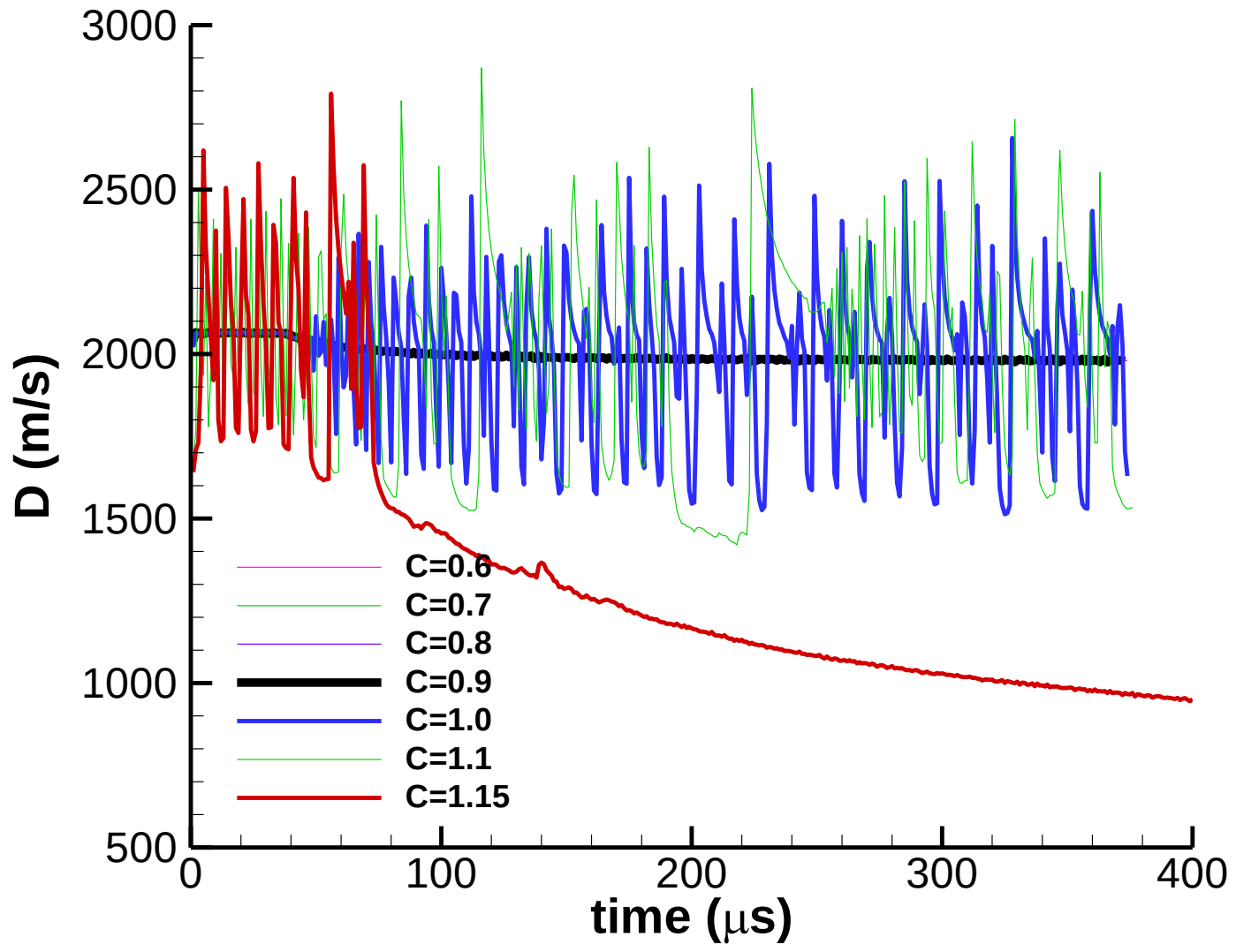

(a) detonation velocity

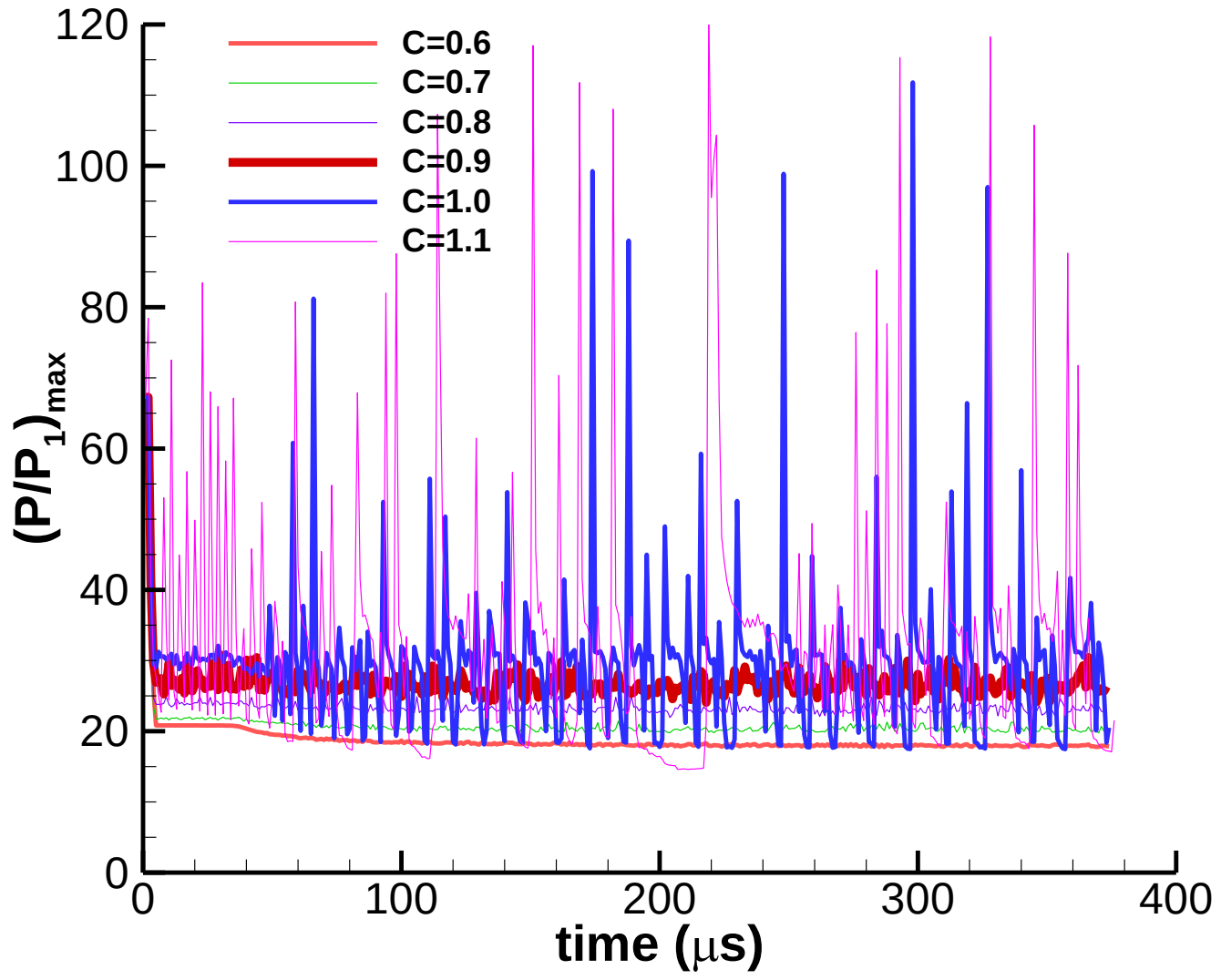

(b) Maximum pressure ratio

Fig.1 Detonation velocity and Maximum pressure ratio under different activation energy

The corresponding maximum pressure ratio on the detonation front are plotted in Fig.1(b). It can be seen from Fig.1(b) that, in the stable region for the parameter *C* from 0.6 to 0.9, the pressure spike is almost a constant, but its value increases as the activation energy increases. In the unstable region for the parameter *C* from 1.0 to 1.1, the maximum pressure oscillates very violently between 15atm and 120atm. The lowest pressure of unstable detonation is about 15atm, which is equal to the theoretical value of C-J detonation of stoichiometric $H_2$-air mixture at initial conditions of 1atm and 300K.

**3.2 the stable detonation (C=0.6-0.9)**

The profiles of convective flux Γ, kinetic energy ΔK, heat release ΔQ, and the algebra summation of these three fluxes of stable detonation are plotted in Fig.2. In the following discussion, for convenience, the summits of heat release and convective flux are labeled as Q and Γ, the bottom of the kinetic energy valley is labeled as K, and the peak of the algebra summation of these three fluxes is labeled as S in the figures, respectively.

It can be seen that, for the case of C=0.6, these four peaks are at the same grid point. The heat release is coupled with the shock wave tightly. The kinetic energy ΔK is negative and its absolute value is smaller. The algebra summation of these three fluxes is positive. With the increase of activation energy, for the case of C=0.7, the point Γ and the point K are at the same grid point of x=206.67mm, but the absolute value of K becomes stronger. The heat release summit point Q and the summation peak S are at x=206.665mm, which is one grid point behind the summit point Γ of convective flux. In addition, a negative convective flux peak point $S_1$ begins to appear, which is a rarefaction wave ($S_1 \approx$ -0.2MJ/m$^3$), and reduces the internal energy per unit volume of detonation products.

For the case of C=0.8, the point Q is still one grid point behind the convective flux summit point Γ, but the strength of the rarefaction wave of negative convective flux point $S_1$ becomes stronger ($S_1 \approx$ -0.7MJ/m$^3$). For the case of C=0.9, the distance between point S and point Γ becomes wider. The heat release peak point Q is two grid points behind the peak point Γ of shock wave. The negative convective peak $S_1$ becomes more stronger ($S_1 \approx$ -1.8MJ/m$^3$) than that of case C=0.8. But it is still two grid points behind the heat release peak Q, which means that the heat release is not influenced by the rarefaction wave significantly. For the stable detonation, the flame is coupled with the shock wave and the detonation propagates self-sustainably.

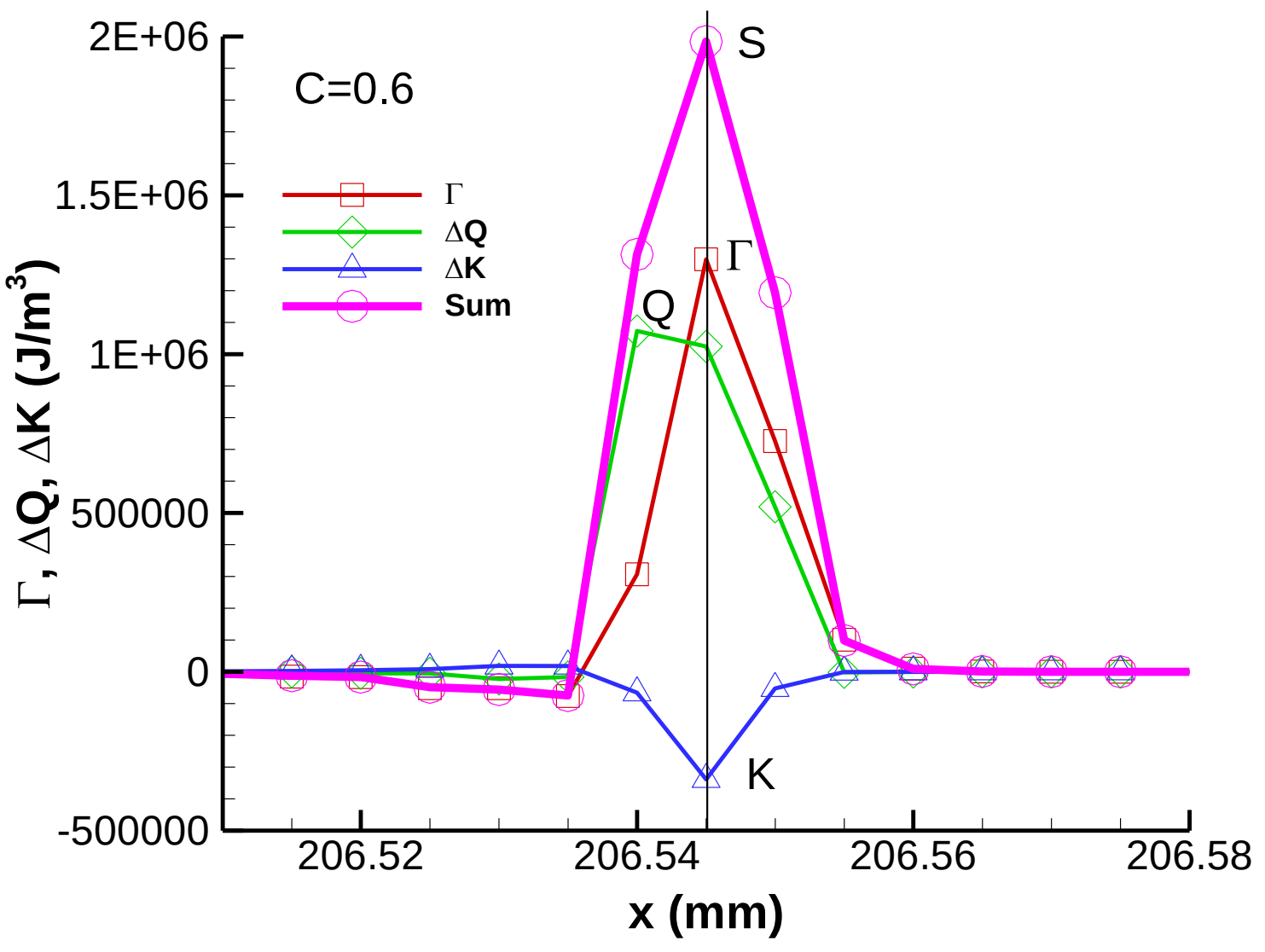

(a) C=0.6

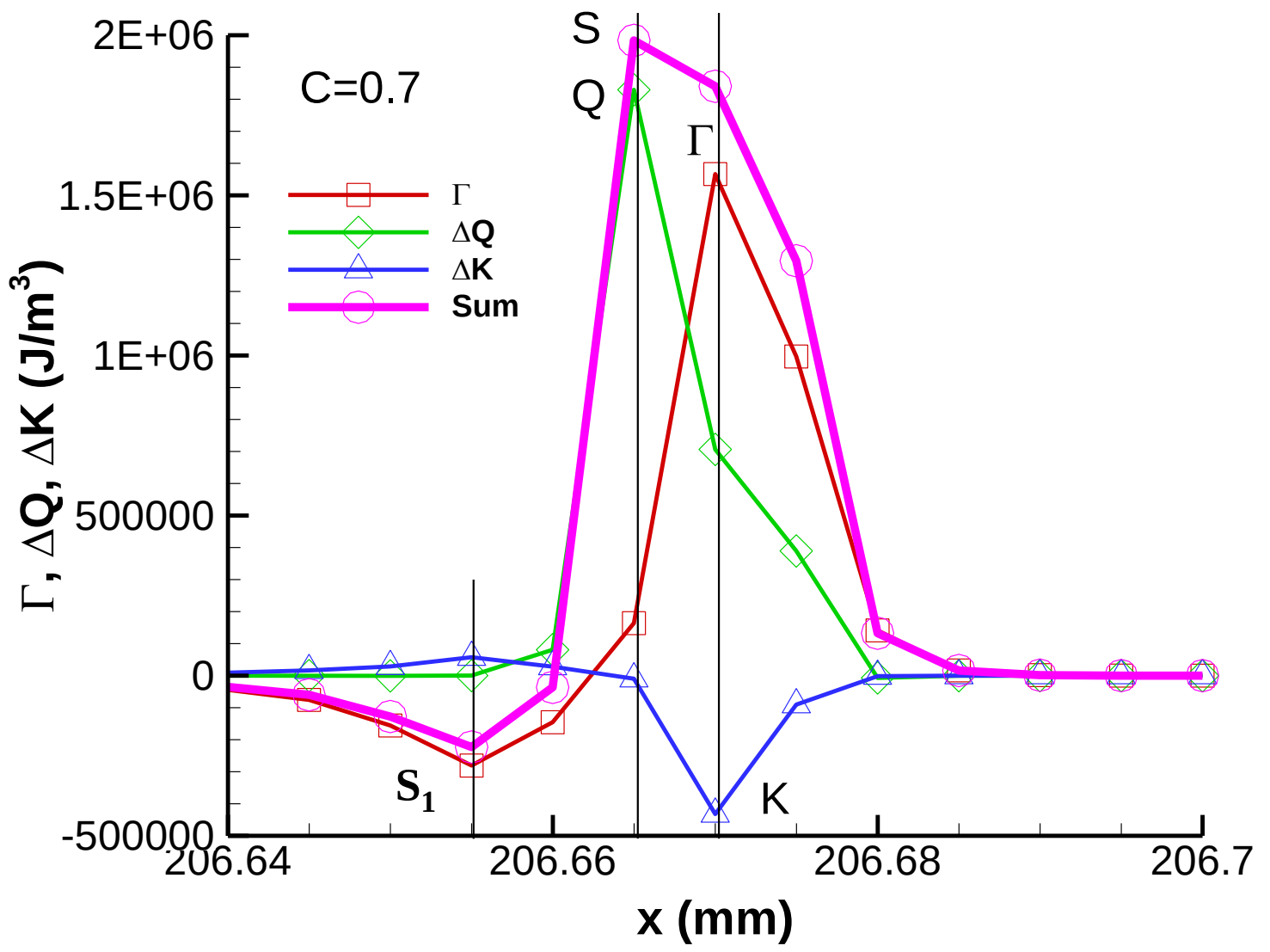

(b) C=0.7

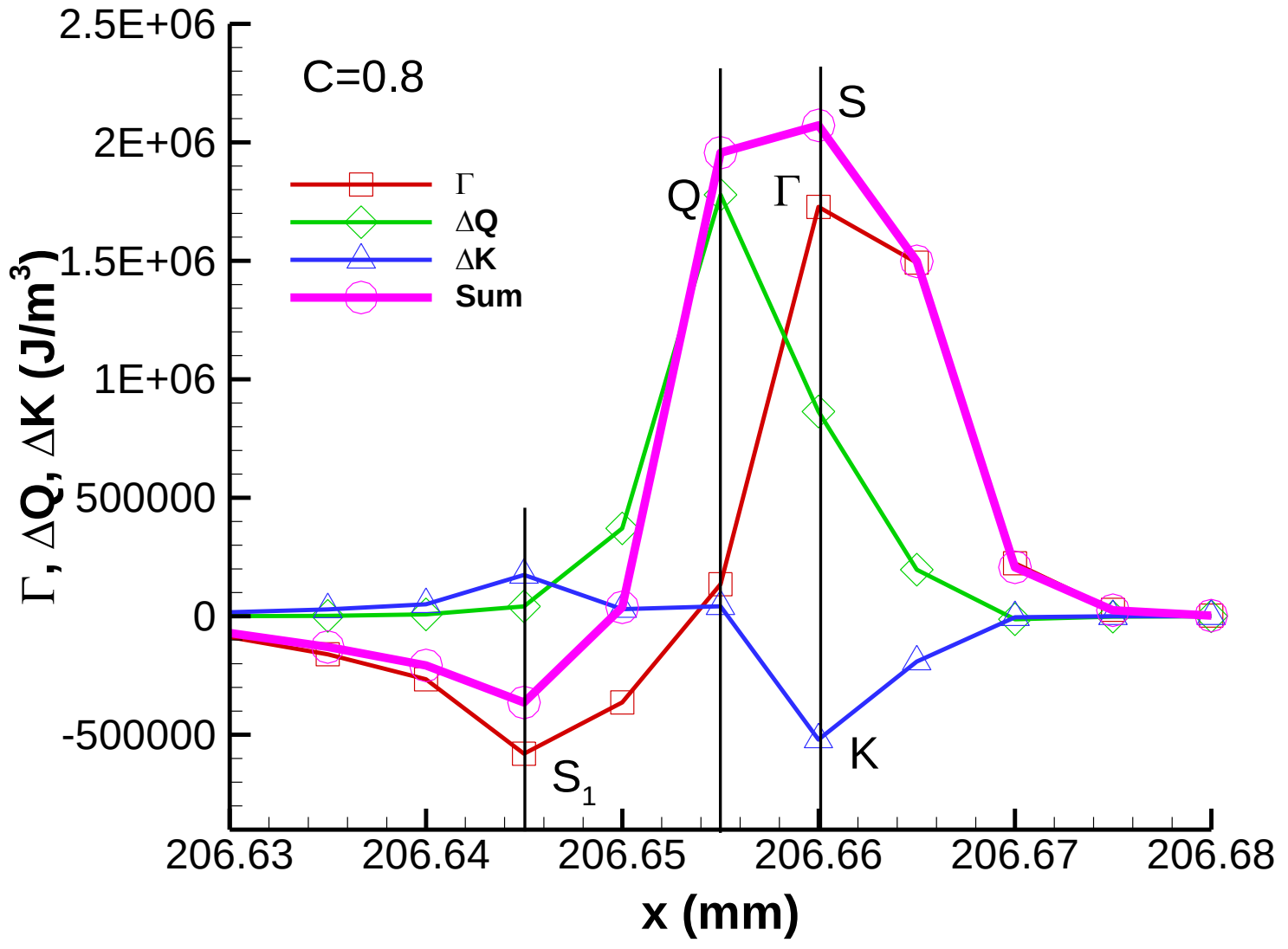

(c) C=0.8

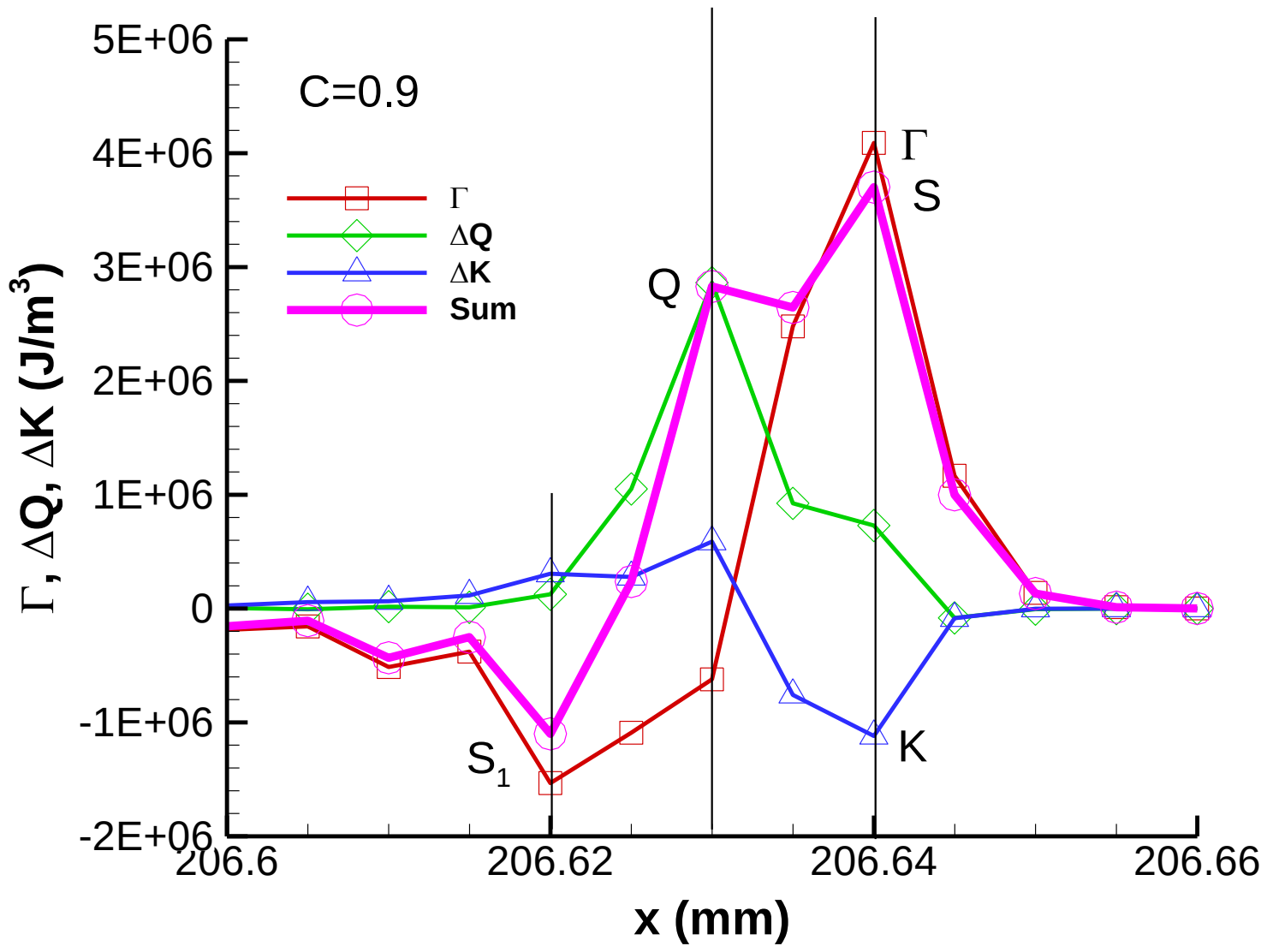

(d) C=0.9

Fig.2 the profiles of convective flux of stable detonation

The corresponding profiles of pressure, the pressure gain ΔP at one time step defined in Eq.(11) and the chemical reaction progress parameter Z of stable detonation are plotted in Fig.3. We can find from Fig.3 that the heat is 100% released on the windward side of the shock wave for the case of C=0.6 with lower activation energy. The heat release produces a pressure gain ΔP of about 4.5atm at the maximum heat release point Q (the vertical thin line) on the detonation front. The pressure on the detonation front is about 18atm, which is equal to the theoretical value of C-J detonation. This case can be referred to constant volume combustion because there is no rarefaction wave on the detonation front.

With the increase of activation energy, for the case of C=0.7, not all heat is released on the windward side of shock wave; some is instead released on the leeward side of shock wave. The maximum pressure gain at the maximum heat release point Q is about 2atm, which is about 2.5atm smaller than that of the case C=0.6. The negative summation peak point $S_1$ generates a negative pressure gain ($\Delta P<0$) at x=206.655mm, which decreases the pressure of detonation products. For the case of C=0.8, the pressure gain at the heat release point Q is about 1.5atm. The rarefaction wave generates a negative pressure gain of about $\Delta P$=-1atm at x=206.645mm. For the case of C=0.9, this condition becomes severe. The positive pressure again at point Q is about 3atm. But the negative pressure again by the rarefaction wave is about -4atm at x=206.625mm, which is only one grid point behind the heat release point Q. In summary, the stable detonation wave can obtain positive pressure gain by heat release and the rarefaction wave is always behind the flame. Therefore, it can be also referred to as pressure gain combustion, and the heat lease can support the propagation of shock wave.

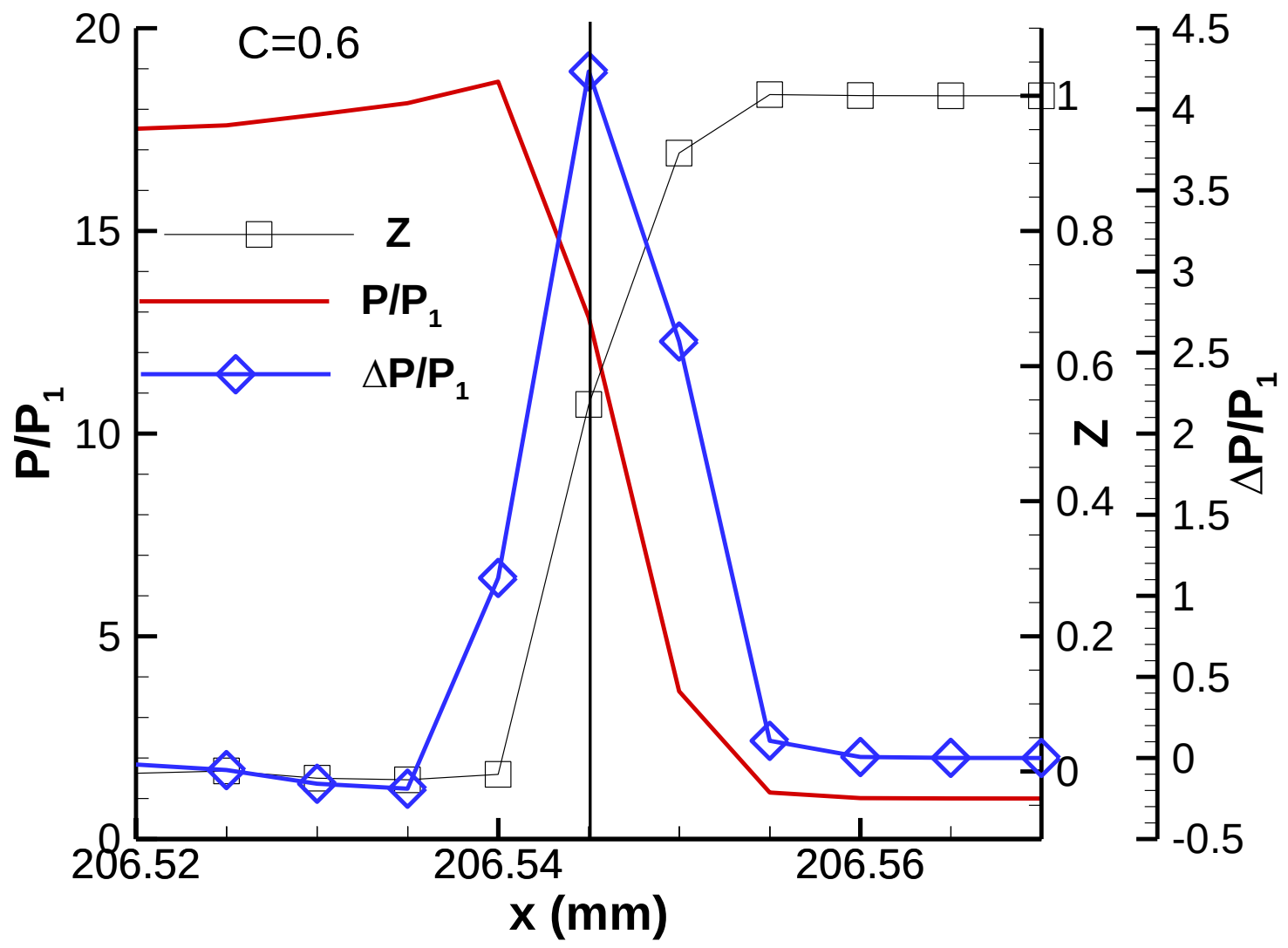

(a) C=0.6

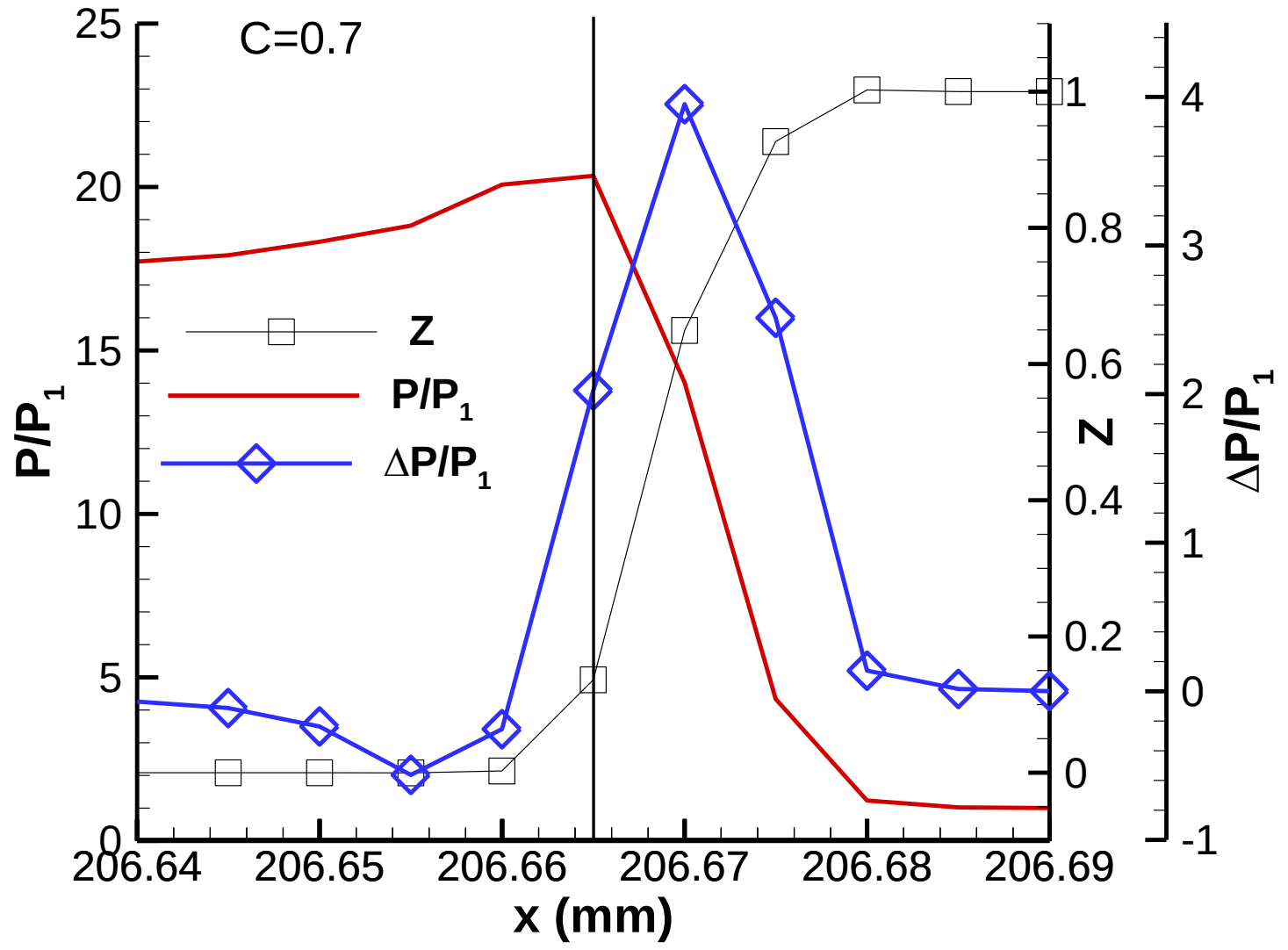

(b) C=0.7

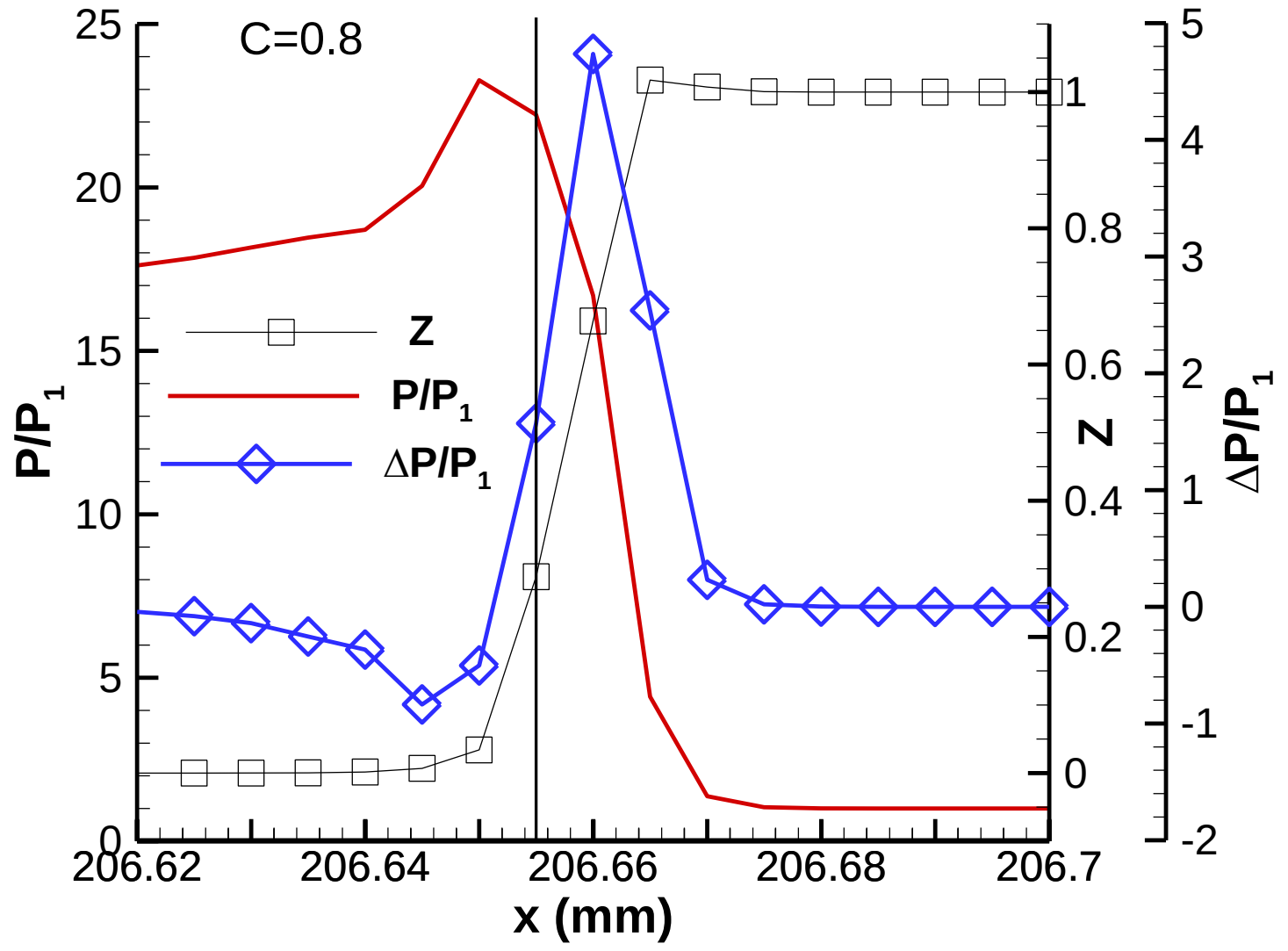

(c) C=0.8

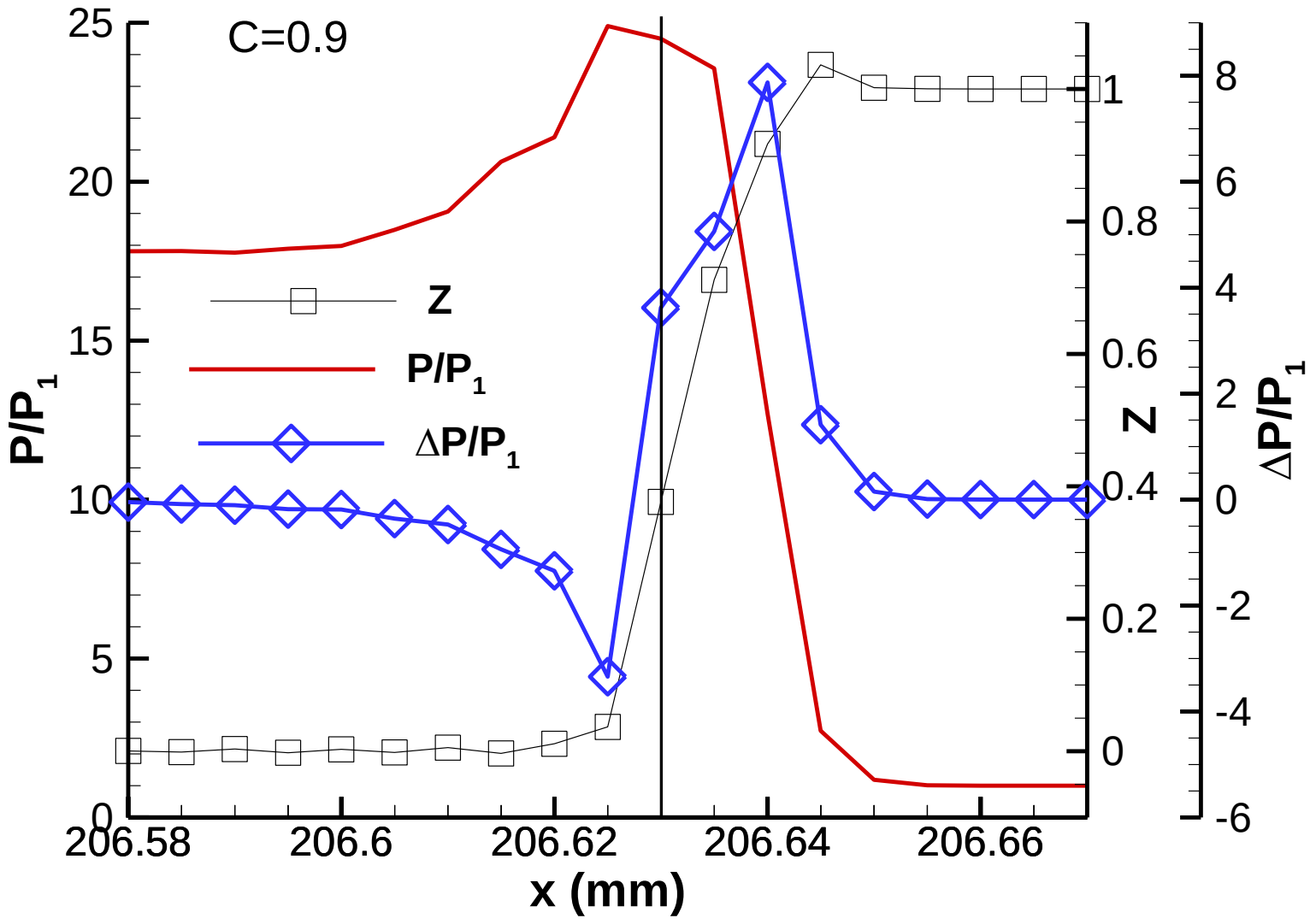

(d) C=0.9

Fig.3 the profiles of pressure and pressure gain of stable detonation

### 3.3 the periodic detonation (C=1.0)

The detonation wave becomes periodically unstable when the activation energy increases to C=1.0. The detonation velocity and maximum pressure ratio of the periodic detonation are plotted in Fig.4. It looks like that the detonation has the regular and periodic oscillation behavior. The enlarged local pressure profiles from 100μs to 106μs are plotted in Fig.5. It can be seen from Fig.5 that the detonation wave is quenched at t=103μs and 104μs, but is reignited at t=105μs.

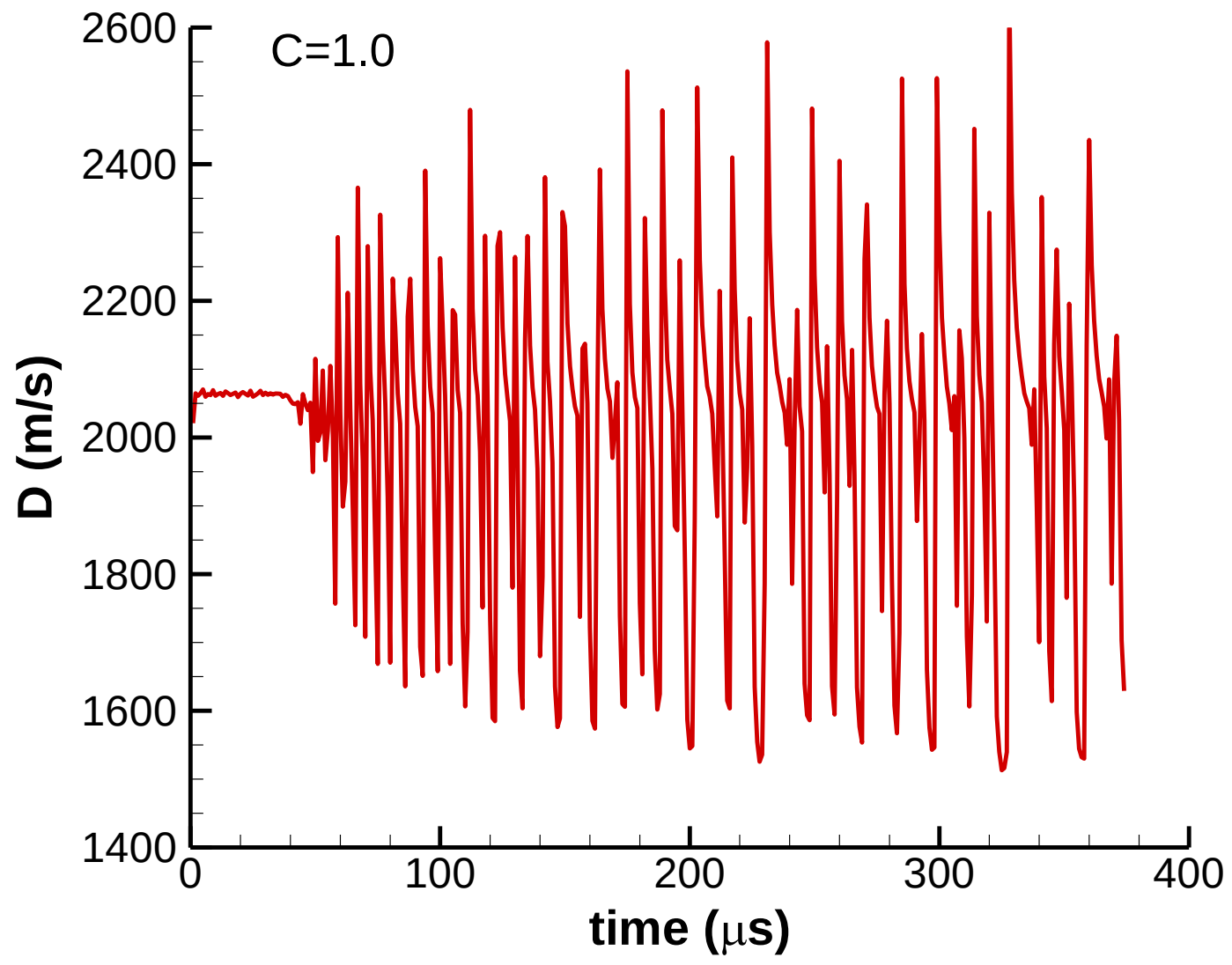

(a) detonation velocity

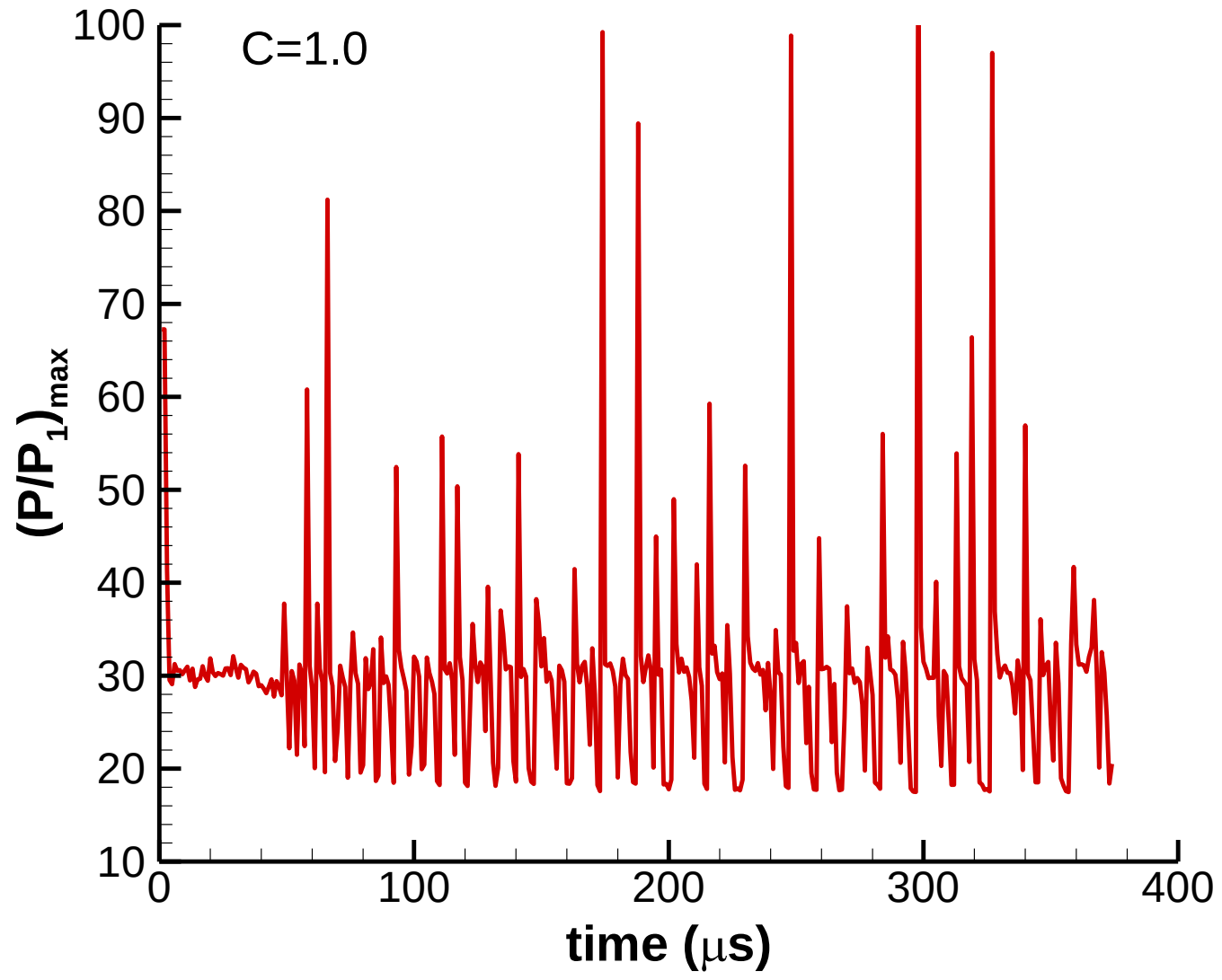

(b) maximum pressure ratio

Fig.4 the detonation velocity and maximum pressure of periodic detonation

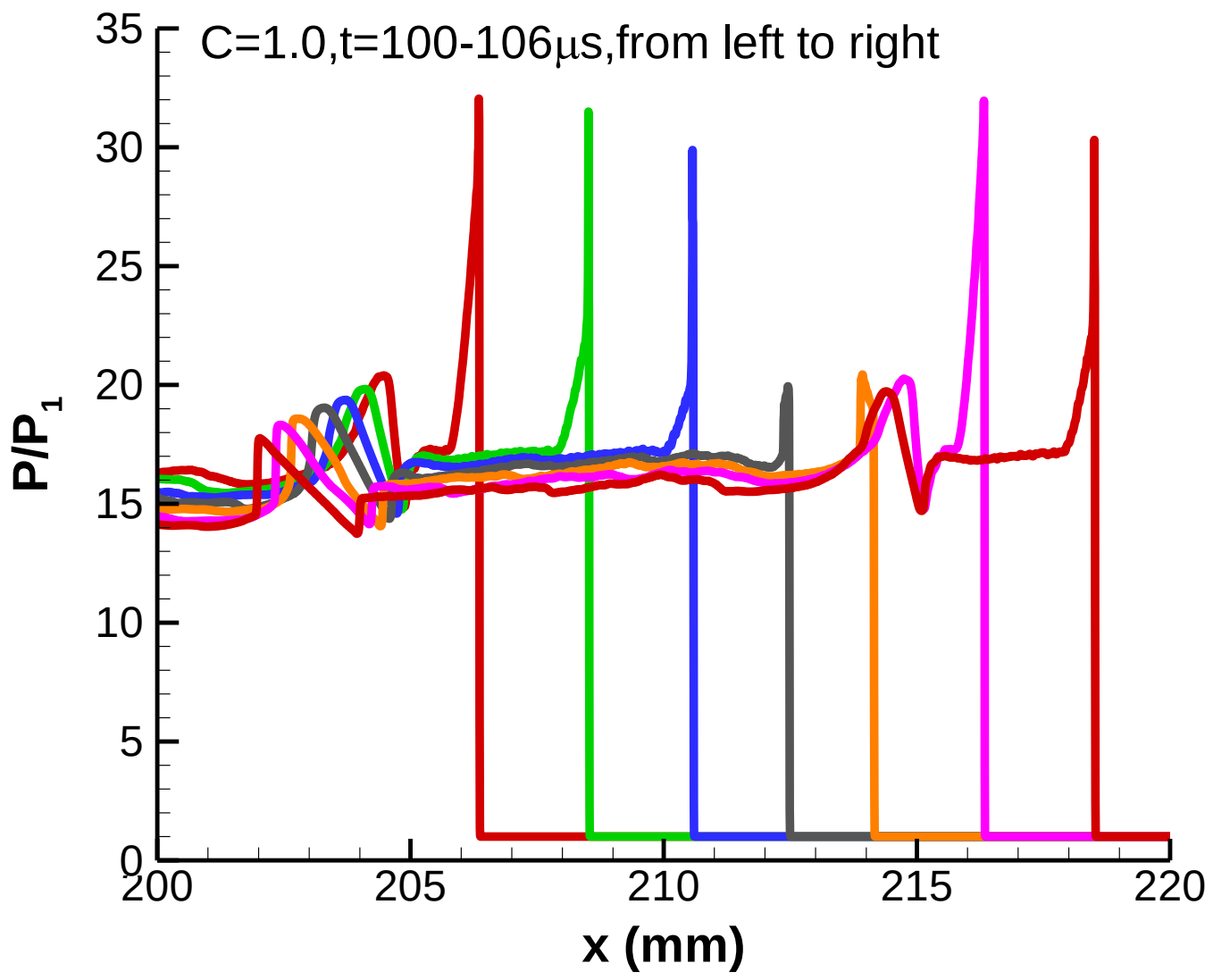

Fig.5 the pressure profiles of periodic detonation

This periodic behavior can be explained by the flux analysis mythology. The convective fluxes from 100μs to 105μs are shown in Fig.6. At t=100μs, the heat release is coupled with the shock wave and there are two peaks of heat release. The negative convective flux peak point $\Gamma_1$ coincides with the second heat release summit point $Q_1$ at x=206.35mm and counteracts the heat release there. Accordingly, the pressure and temperature there are decreased and the chemical reaction rate becomes slower.

At t= 101μs, the point Q is two grid points behind the shock wave. The negative convective flux point $\Gamma_1$ becomes stronger and is only one grid point behind the heat peak point Q. At t= 102μs, the negative convective flux point $\Gamma_1$ moves forward and is at the same grid point with the point Q at x=210.57mm. The total summation of these three fluxes is very small, which means that the internal energy of combustion products does not increase too much although combustion takes place there. At t=103μs and 104μs, the flame fully decouples from the shock wave. The flame coincides with the negative convective flux and the summation is almost zero. The combustion becomes constant pressure combustion in a sense at these two instants. The distance between the flame and the shock wave becomes wider as time goes by. However, at t=105μs the flame is recoupled with the shock wave again by autoignition. The detonation wave propagates periodically following this mechanism.

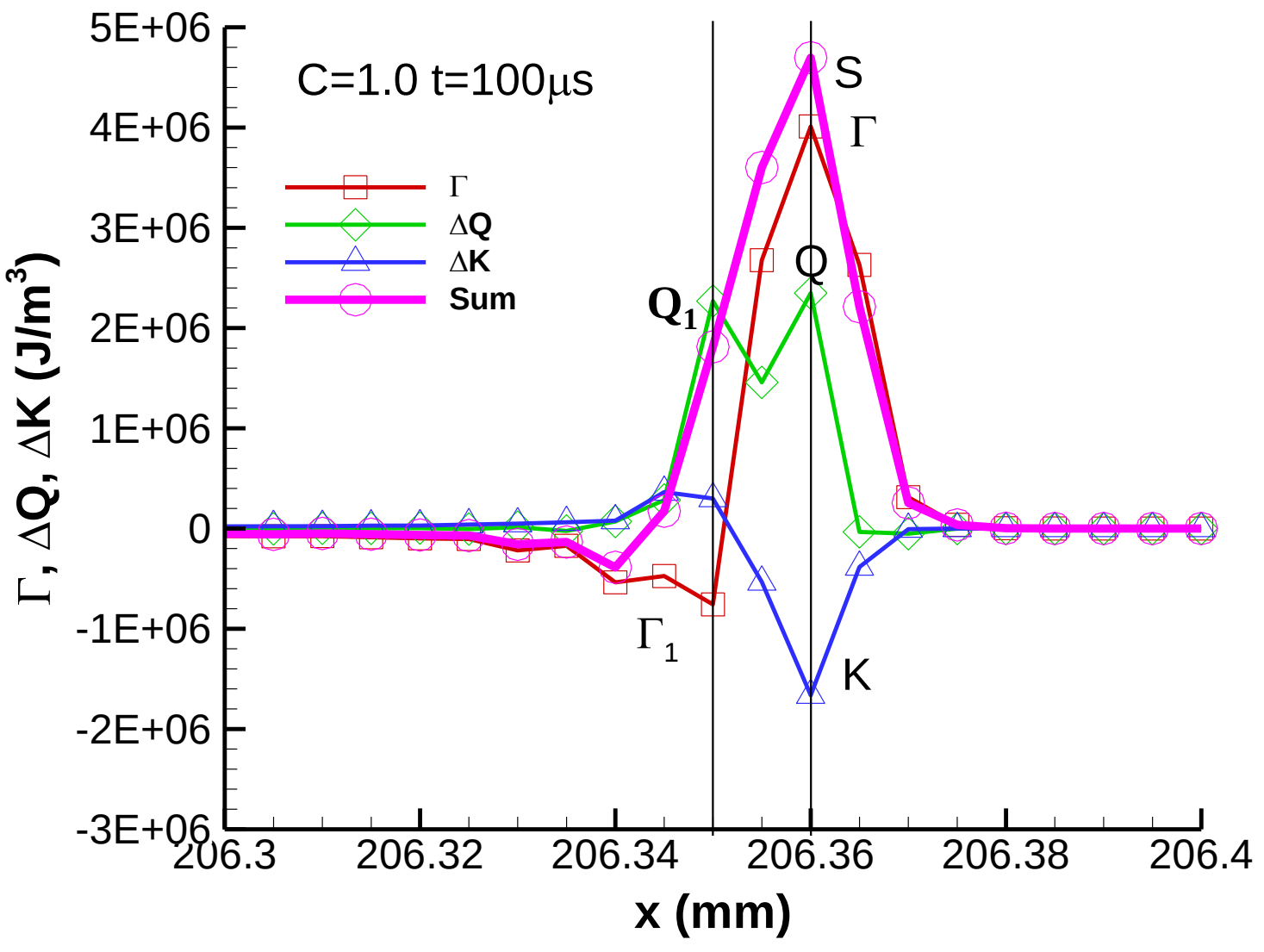

(a) t=100μs

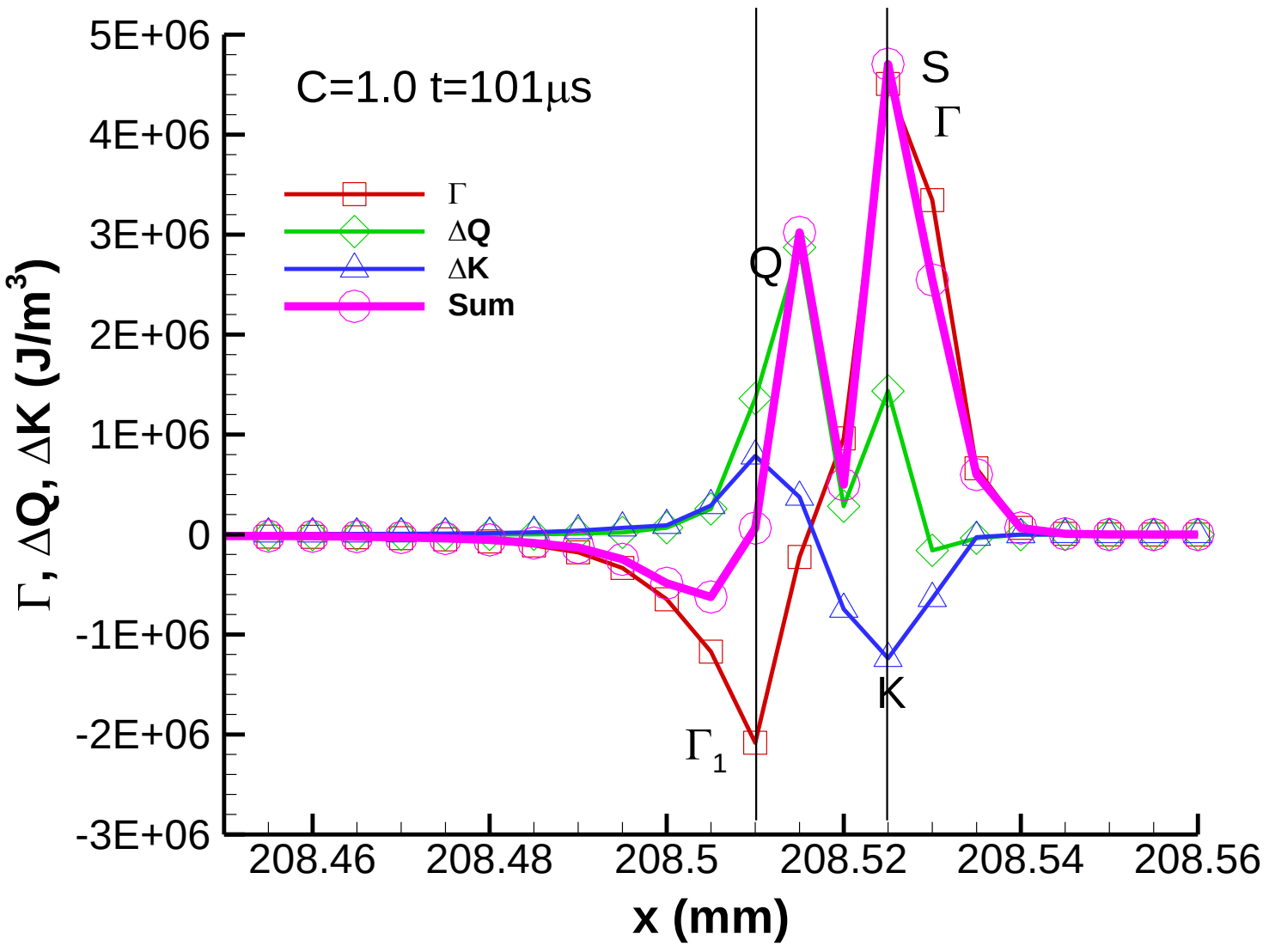

(b) t=101μs

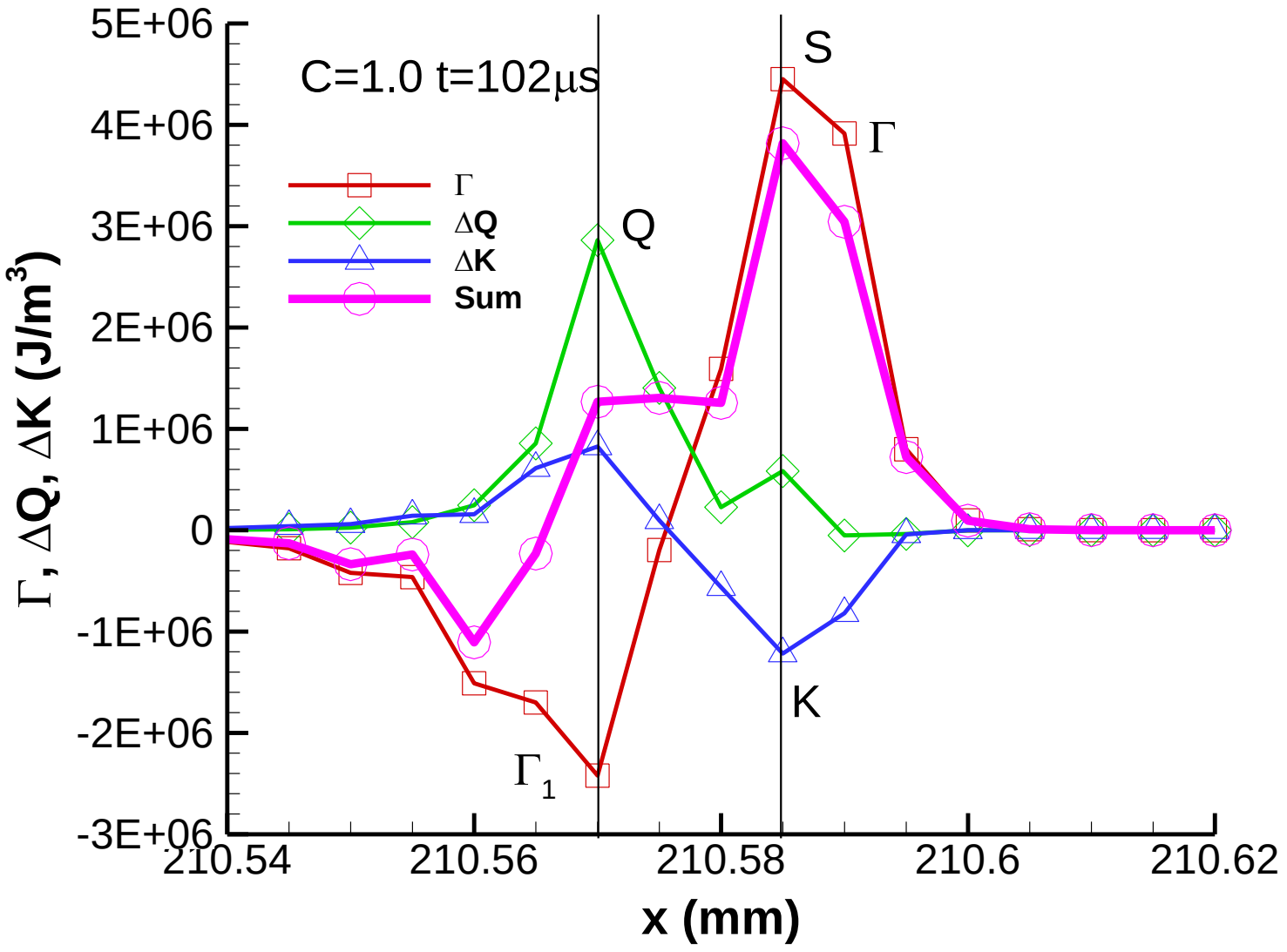

(c) t=102μs

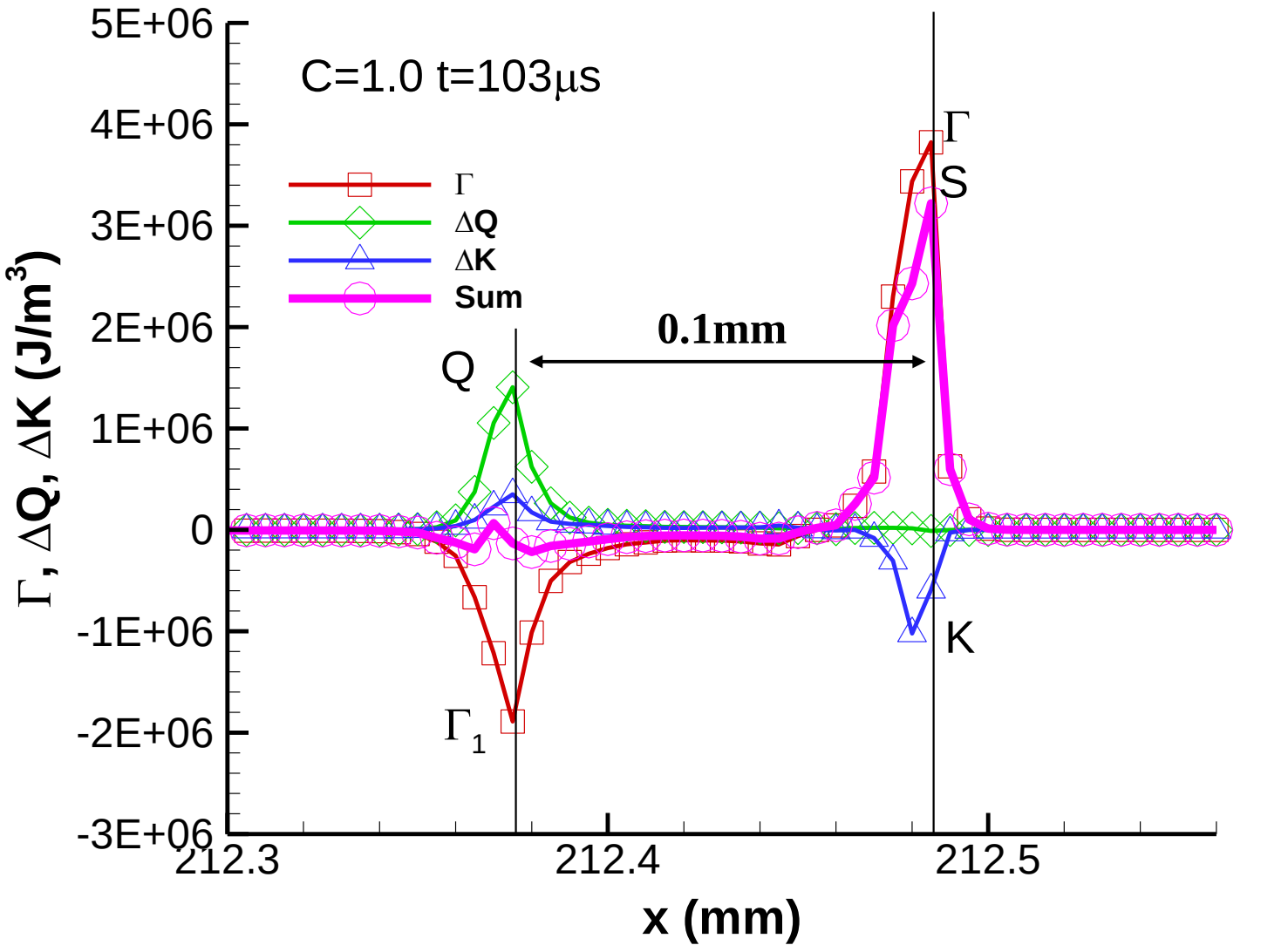

(d) t=103μs

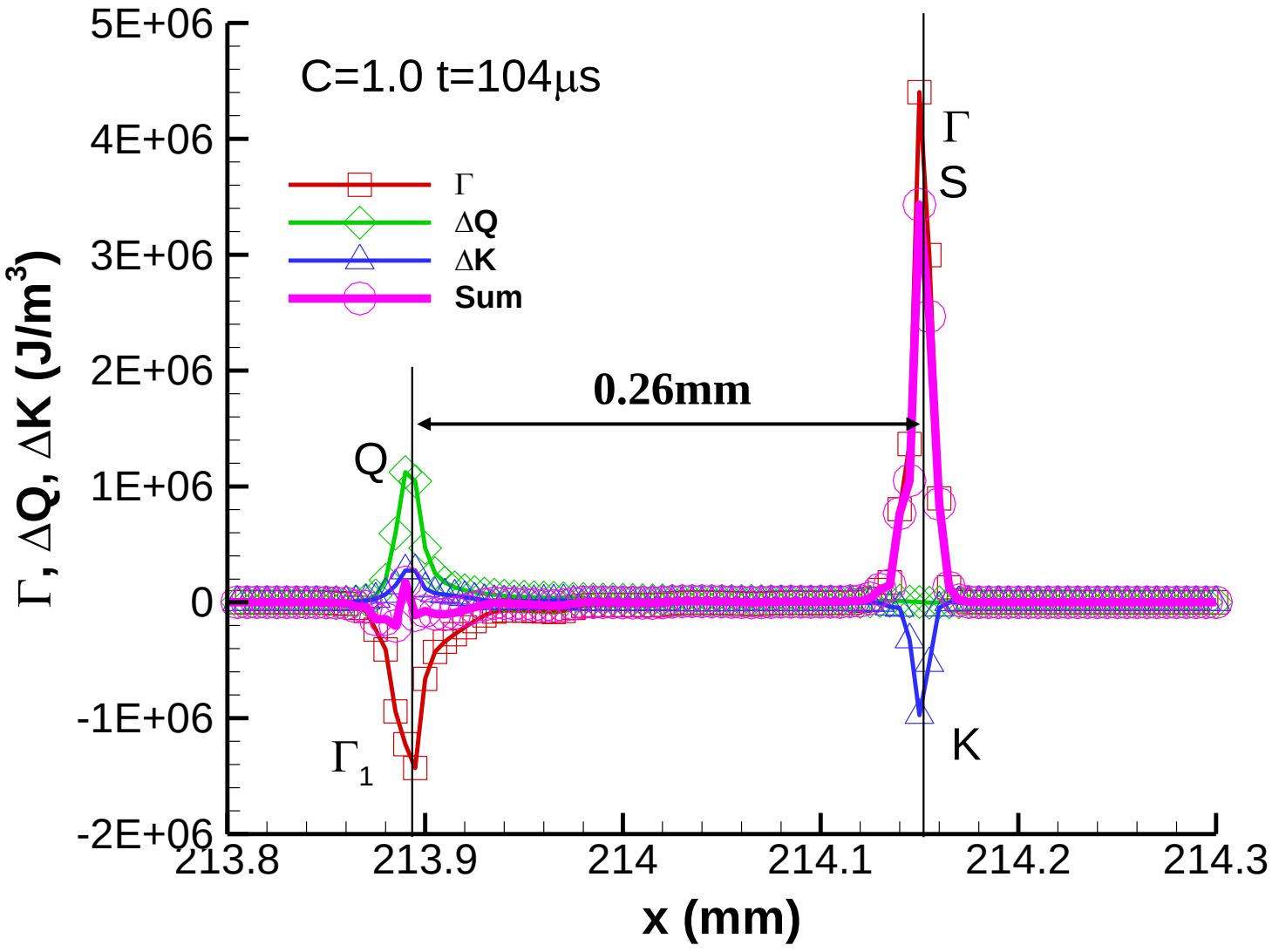

(e) t=104μs

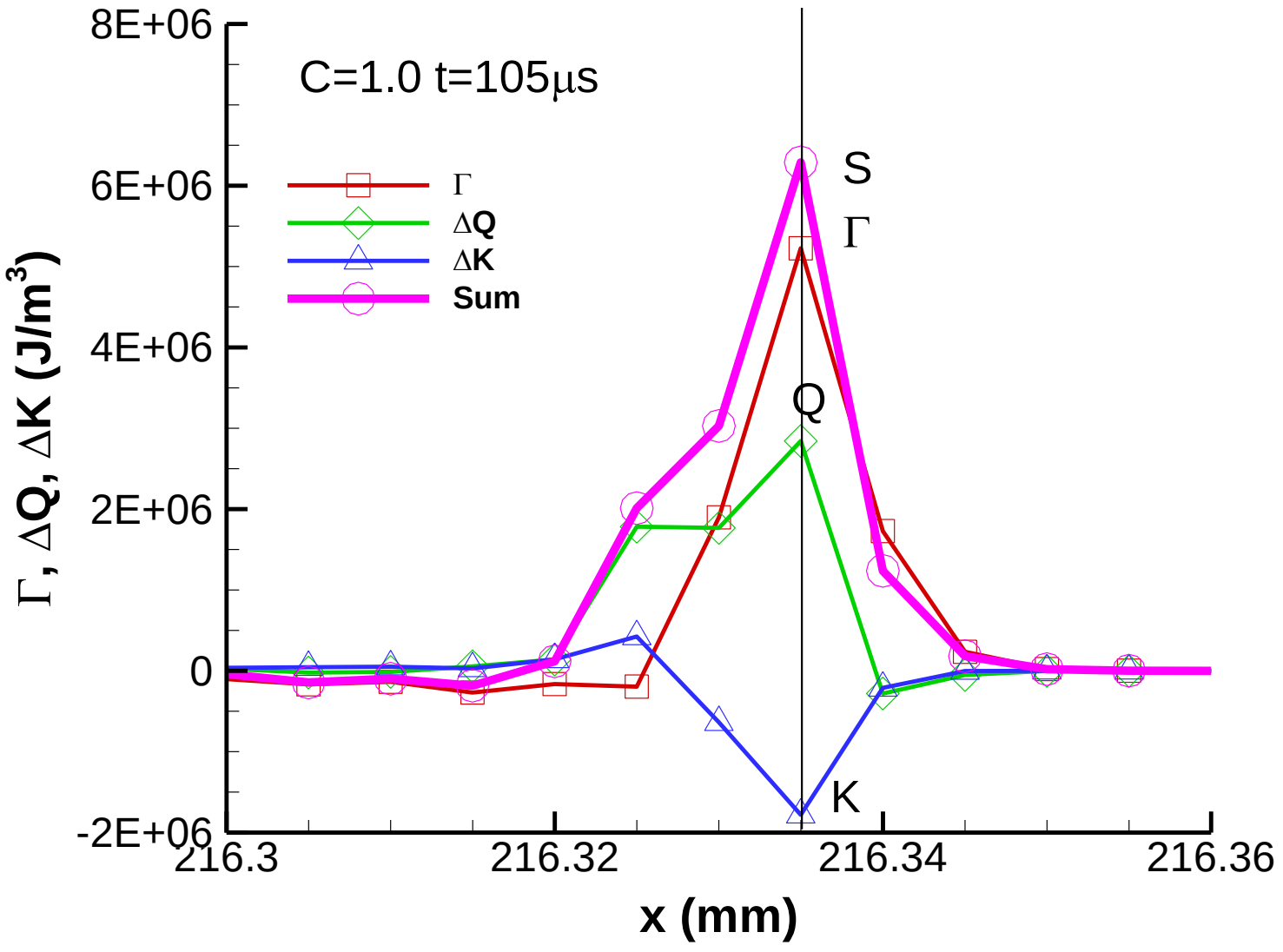

(f) t=105μs

Fig.6 the profiles of convective flux of periodic detonation

### 3.4 Key mechanism of detonation instability

The above discussion indicates that the negative convective flux plays a crucial role in detonation instability. This negative convective flux is generated by the shaper and steeper enthalpy flux $F_E$ at the detonation front, which is defined in Eq.(12). The corresponding enthalpy fluxes of the periodic detonation at different instants are shown Fig.7. We can see from Fig.7 that the gradient of enthalpy flux at detonation front is very large.

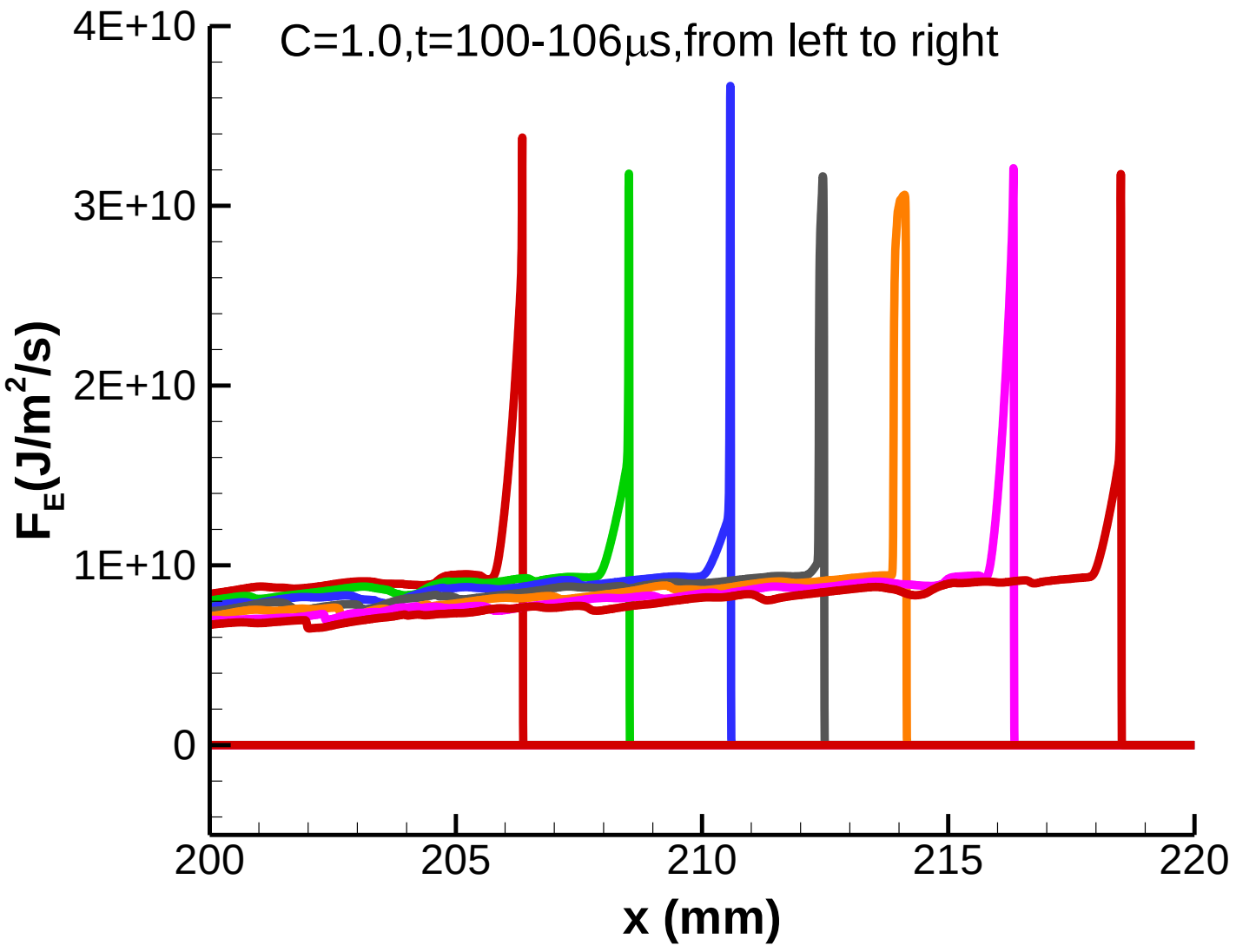

Fig.7 The enthalpy fluxes of periodic detonation

The reason is that the total enthalpy on the detonation front varies very significantly once the flame is decoupled from the shock wave during detonation propagation. The theoretical total enthalpy on theoretical C-J detonation point and a Ma4.85 normal shock wave front (gaseq results) are listed in Table 2. The Ma4.85 is the Mach number of C-J detonation. It can be seen from Table 2 that the total enthalpy of the shock wave front is almost more than 4 times larger than that of the C-J detonation front. This means that the enthalpy flux becomes very sharper and steeper once the flame is decoupled from the shock wave. Consequently, the sharper and steeper enthalpy flux produces a very stronger negative convective flux, which reduces the total energy per unit volume of detonation products.

**Table 2. Theoretical parameters on C-J detonation and shock front**

| Parameters | C-J detonation | Ma4.85 Shock | Ratio |
|---|---|---|---|
| Density (kg/m$^3$) | 1.49 | 4.58 | 0.32 |
| P(atm) | 15 | 27.9 | 0.53 |
| T(K) | 2956 | 1548 | 1.9 |
| Gas velocity(m/s) | 849 | 1612 | 0.52 |
| Total enthalpy (MJ/m$^3$) | 8.19 | 31.67 | 0.25 |

The density profiles and gas velocity profiles of detontion under different activation energy are drawn in Fig.8(a) and Fig.8(b), respectively. The maximum density and maximum gas velocity on the detonation front increase as the increase of activation energy. Their values are between the theoretical value of C-J detonation and the Ma4.85 normal shock wave theoretical value, which are consitent with the data in Table 2. These numerical resutls confirm the key role of convective flux on the detonation instablity.

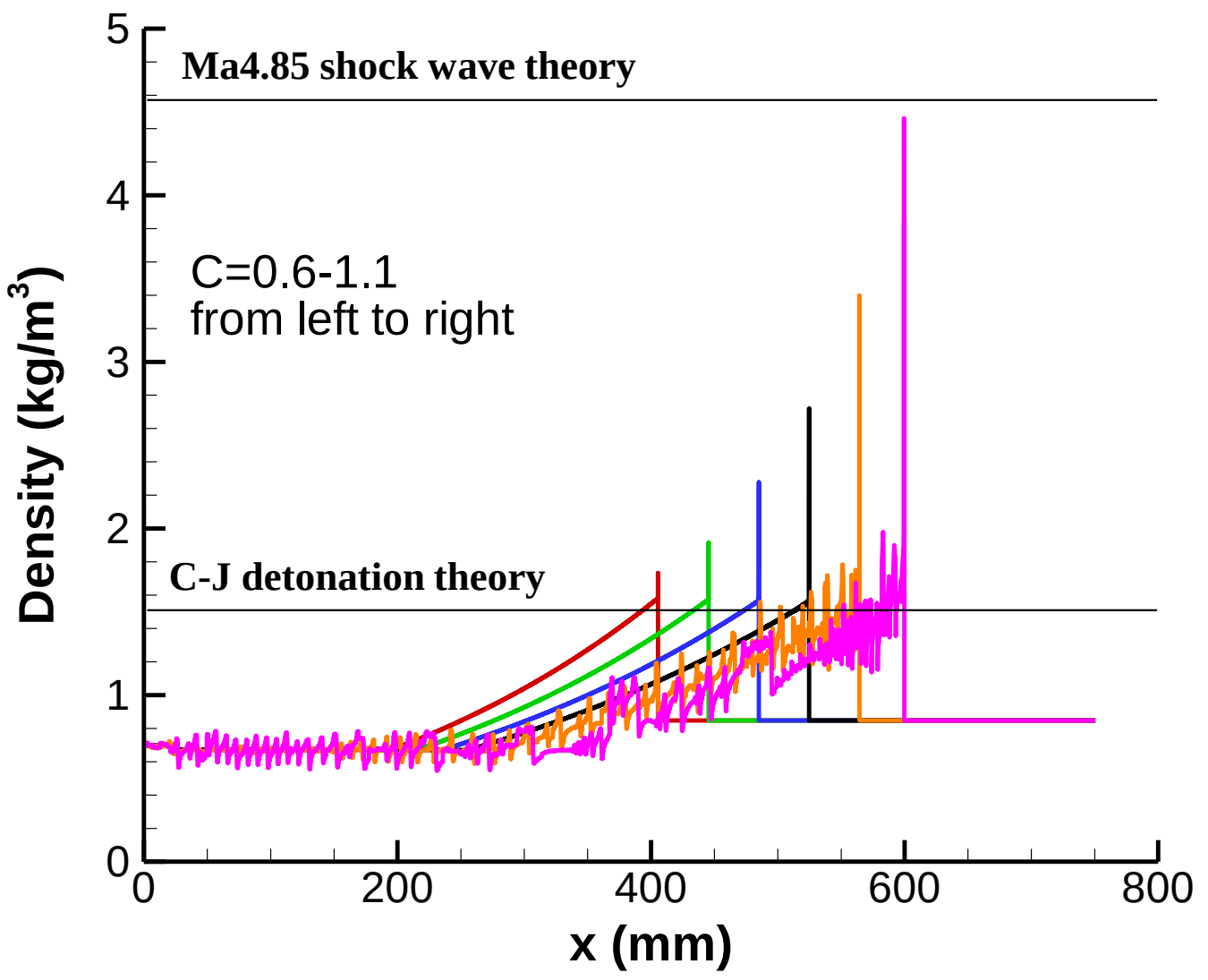

(a) the density profile

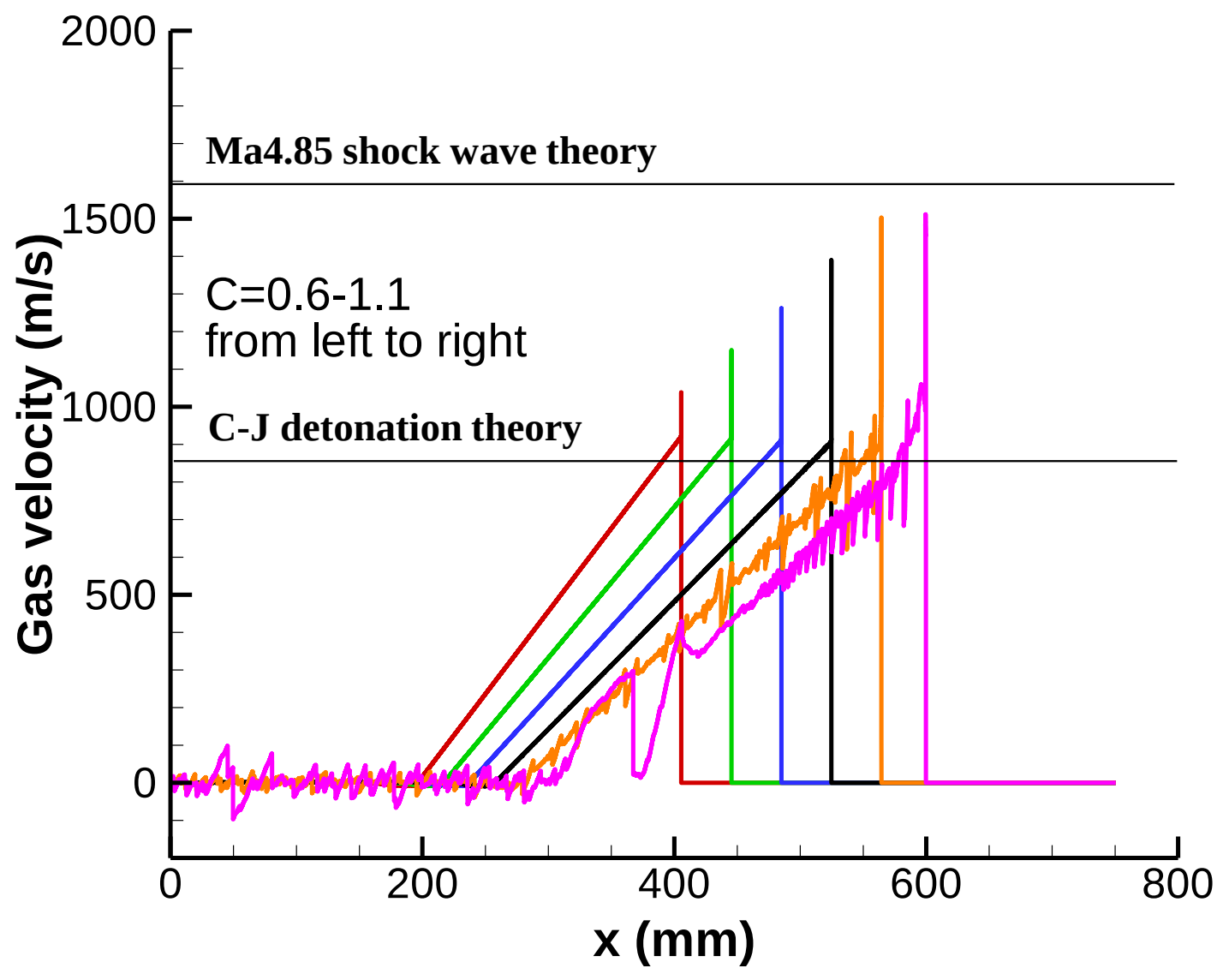

(b) the gas velocity profile

Fig.8 The denstiy and gas velocity profiles under different activation energy

The numerical resutls of the enlarged enthalpy flux $F_E$ and the chemical reaction progress paramter Z at t=103μs are plotted in Fig.9. It can be found that the combustion induces a significant drop of enthalpy flux. In other words, the system transition from reactants to detonation products leads to the system transition of enthalpy flux from reactants to combustion products synchronously. Therefore, we can find that there is always a negative convective flux accompaning the heat release once the flame is decoupled from the shock wave. They are of “mutual generation and restriction” by using the expression of the traditional Chinese Wu-Xing (Five Elements) philosophy. It is the intrinsic nature of the one-dimensional reactive Euler system.

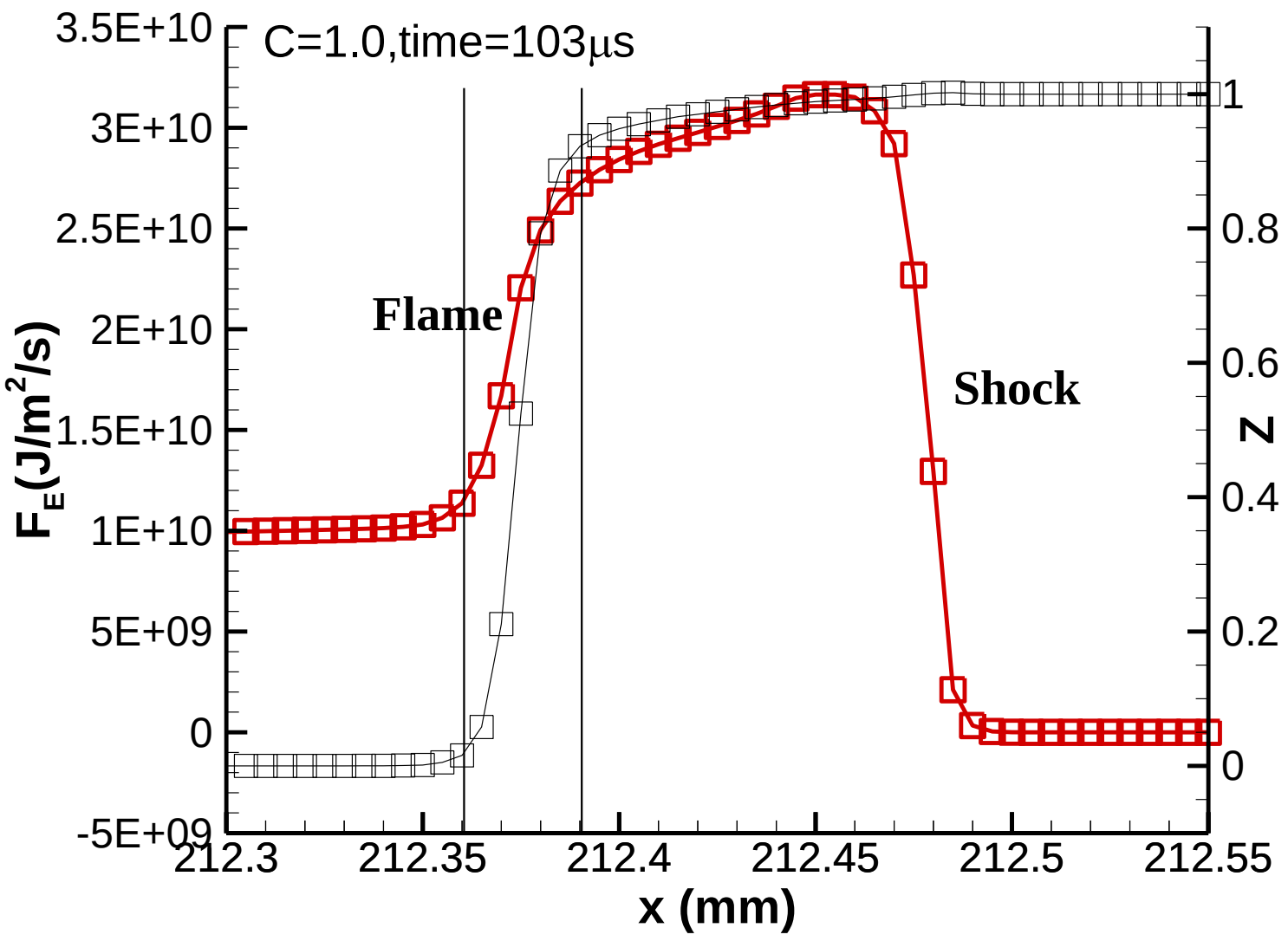

Fig.9 The enlarged enthalpy flux at detonation front

Table 3. Raw CFD data on the periodic detonation front (t=103μs)

| Parameters | $Z_2$=0.000525 ($x_2$=212.3549mm) | $Z_1$=0.922388 ($x_1$=212.3899mm) | Ratio |
|---|---|---|---|
| (ρE+p)u (GJ/m²/s) | 10.643565098 | 27.252431740 | 0.39 |
| (ρE+p) (MJ/m³) | 9.8315187701 | 20.438580481 | 0.48 |
| u (m/s) | 1050.0912067 | 1289.5199184 | 0.81 |
| ρE (MJ/m³) | 8.0955706013 | 18.501844005 | 0.43 |
| $p/p_1$ | 17.128250308 | 19.109388017 | 0.89 |
| E (MJ/kg) | 4.7977229115 | 5.5738917896 | 0.86 |
| ρ (kg/m³) | 1.6873776895 | 3.3193762461 | 0.51 |

The corresponding raw CFD data on the chemical reaction front are presented in Table 3. It can be seen that the enthalpy flux (ρE+p)u of detonation products decreases to be 0.39 times of that of reactants within the flame thickness of 35μm. In the sharp decrease of enthalpy flux (ρE+p)u, the contribution of enthalpy (ρE+p) is larger than the contribution of velocity, which are 0.48 times and 0.81 times, respectively. The variation of energy per unit volume ρE plays important role than pressure in the variation of enthalpy (ρE+p), which are 0.43 times and 0.89 times, respectively. In the energy per unit volume ρE, the density ρ is a very important parameter, which decreases to be 0.51 times of reactants. The variation of specific energy E is smaller, which is 0.86 times of reactants.

### 3.5 the pulsating detonation (C=1.1)

The detonation velocity and maximum pressure ratio of pulsating detonation (C=1.1) are shown in Fig.10. It can be found that the detonation shows the pulsating behavior or chaotic behavior under higher activation energy. The detonation velocity oscillates very violently between 1400m/s and 2900m/s, and the maximum pressure ratio oscillates very violently between 15 and 120. Additionally, it looks like that the detonation is quenched at about t=210μs and the reignition occurs at about t=220μs.

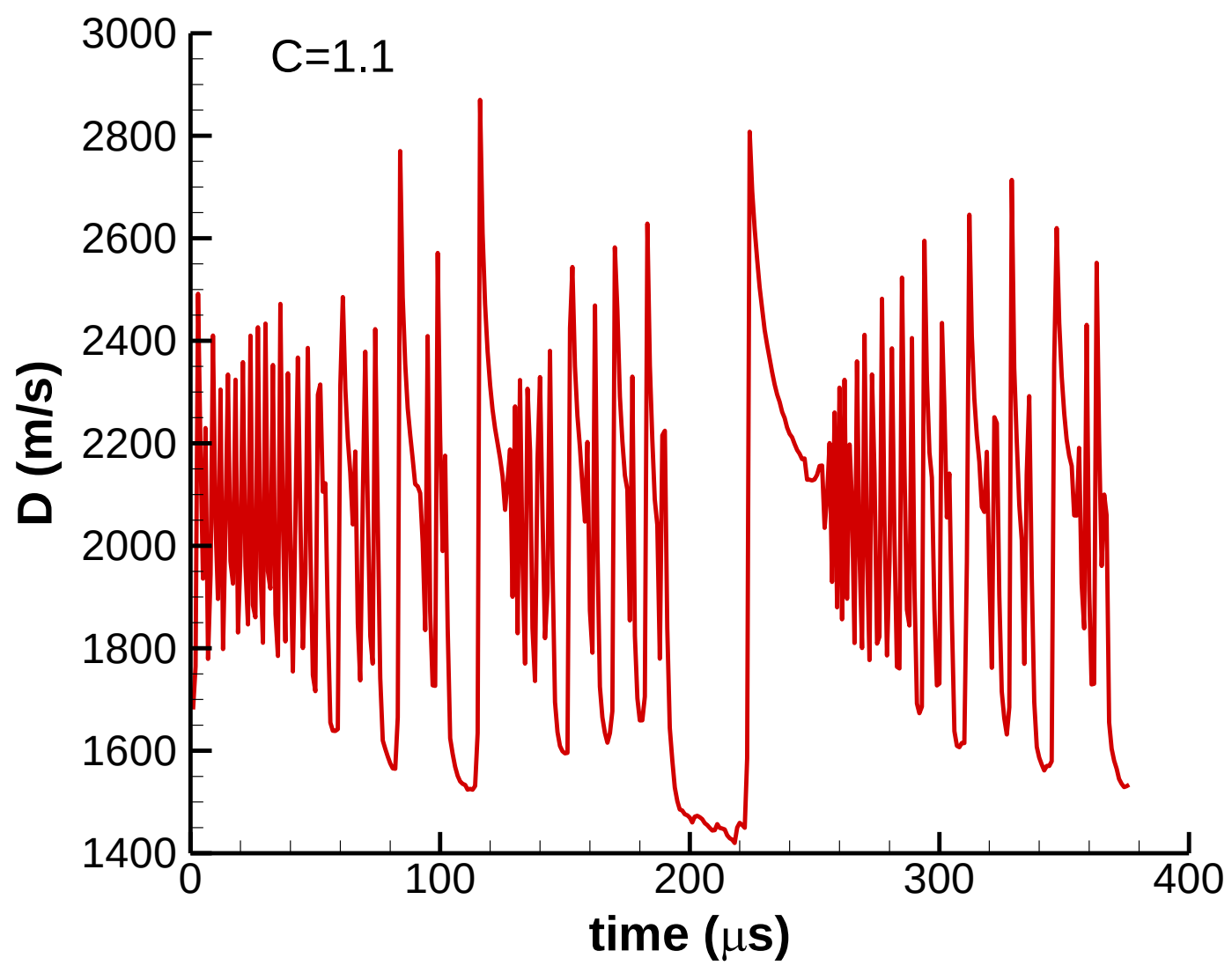

(a) detonation velocity

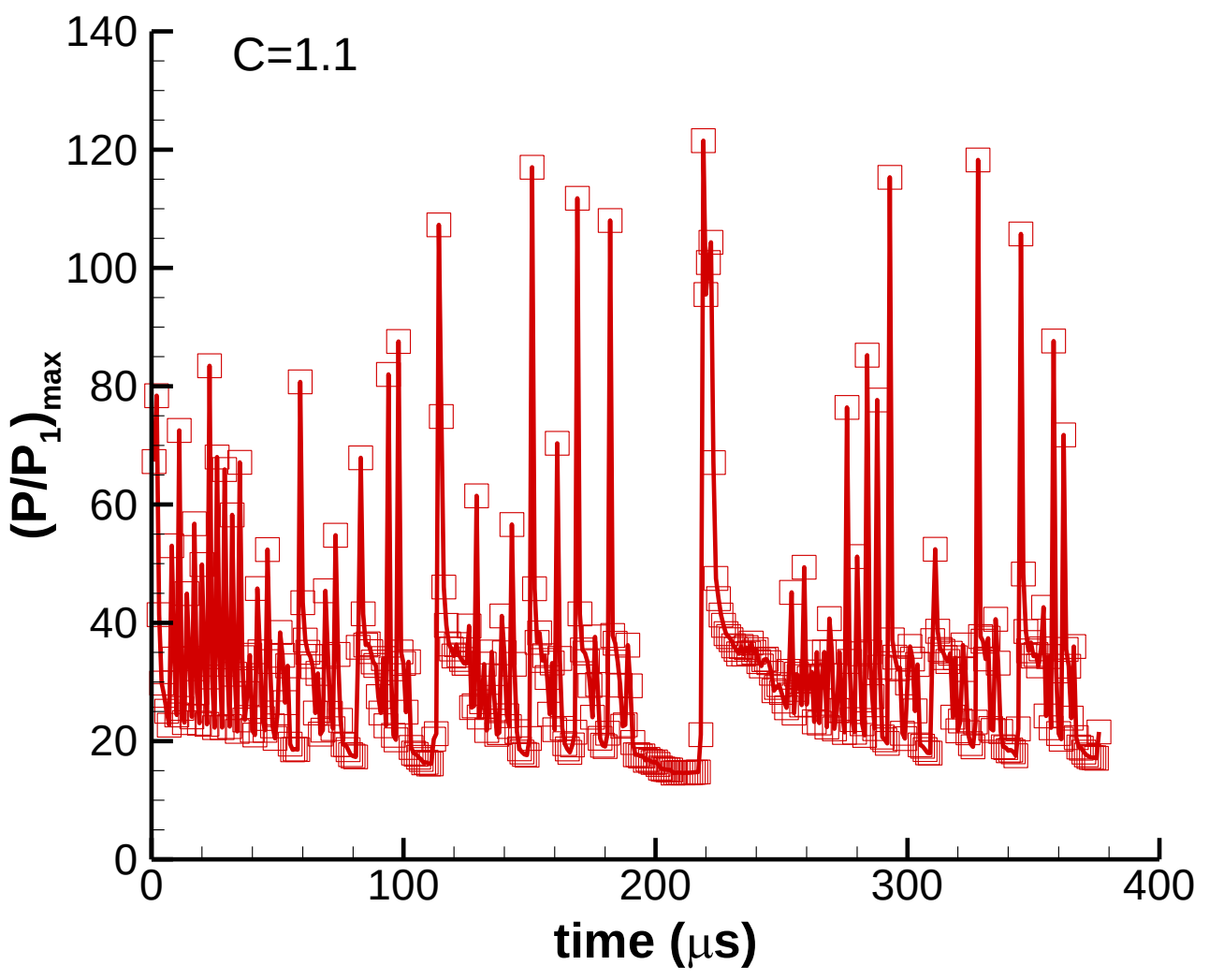

(b) maximum pressure ratio

Fig.10 the velocity and maximum pressure ratio of pulsating detonation

The raw CFD data from t=216μs to 225μs are presented in Table 4. It can be seen that the lowest value of maximum pressure ratio is 14.7 at t=217μs. The maximum pressure ratio reaches the highest value of 121.5 at t=219μs. The lowest detonation velocity is 1420 m/s at t=218μs. The detonation velocity reaches the highest value of 2807.8m/s at t=224μs, which is 5μs later than the maximum pressure ratio. Detonation re-initiation takes place at t=219μs according to the maximum pressure ratio.

**Table 4. Raw CFD data of pulsating detonation**

| Time (μs) | Max Pressure ratio | D (m/s) |
|---|---|---|
| 216.0013826 | 14.74804655 | 1429.370885 |
| 217.0001875 | 14.78441109 | 1426.704896 |
| 218.0000598 | 21.08271153 | 1420.181349 |
| 219.0000025 | 121.5199589 | 1450.083062 |
| 220.0003849 | 95.50619433 | 1459.441851 |
| 221.0002412 | 100.9162851 | 1455.20912 |
| 222.0003638 | 104.3333027 | 1449.822244 |
| 223.0000983 | 67.09051245 | 1585.420866 |
| 224.0008763 | 47.50537086 | 2807.815416 |
| 225.000231 | 44.0902092 | 2691.737026 |

The profiles of flux and pressure gain of pulsating detonation at t=217μs are plotted in Fig.11. We can find from Fig.11(a) that the flame is completely decoupled from the shock front at this instant. The shock wave is at x=426mm, the flame front is at x=418.6mm, and the distance between them is about 8mm. The local enlarged profiles of fluxes are shown in Fig.11(b). The pressure gain at the flame front shown in Fig.11(c) is almost negative, which means that the combustion cannot generate positive pressure to support the shock wave propagation. These results reveal that the negative convective flux is the key mechanism of detonation quenching.

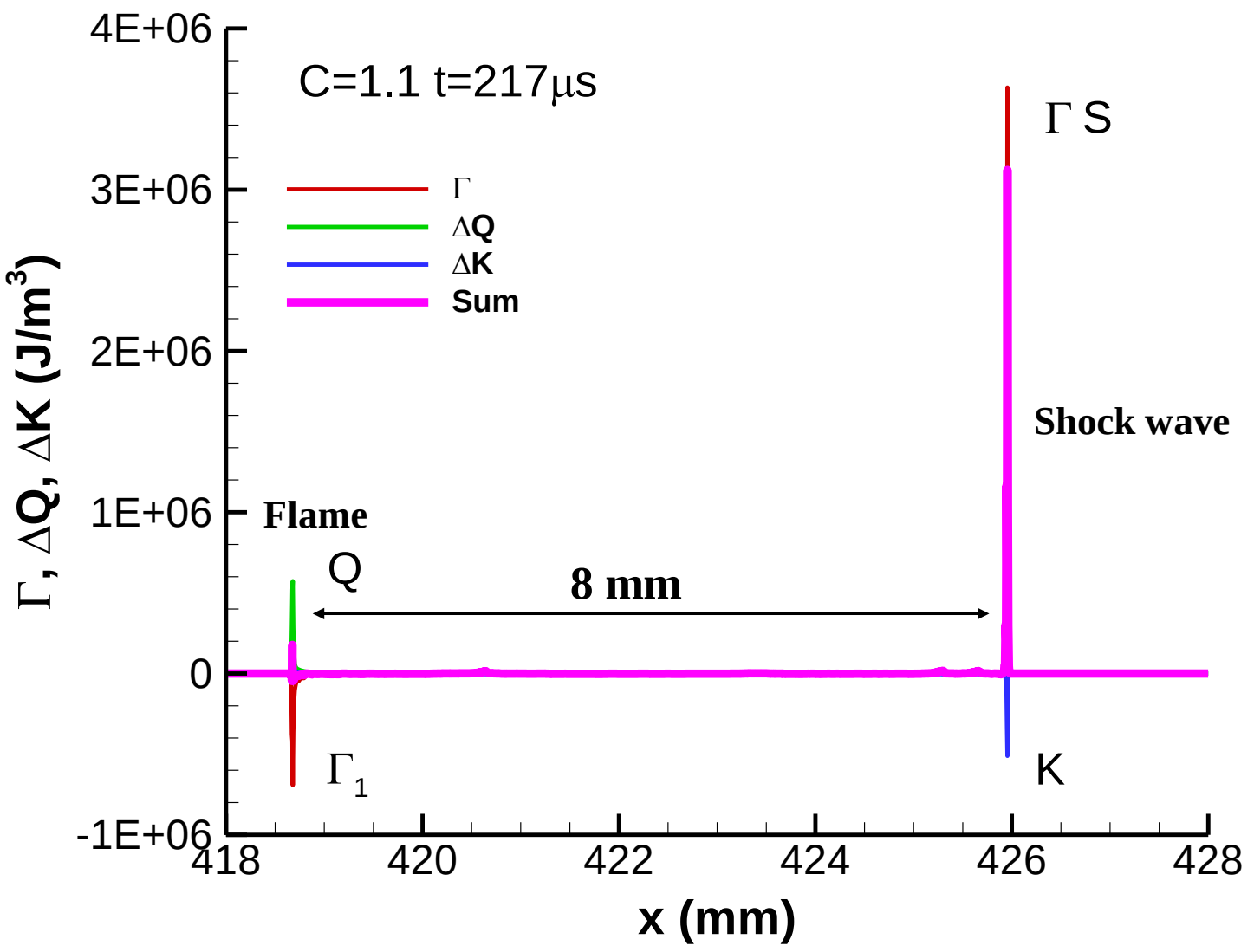

(a) the profiles of flux

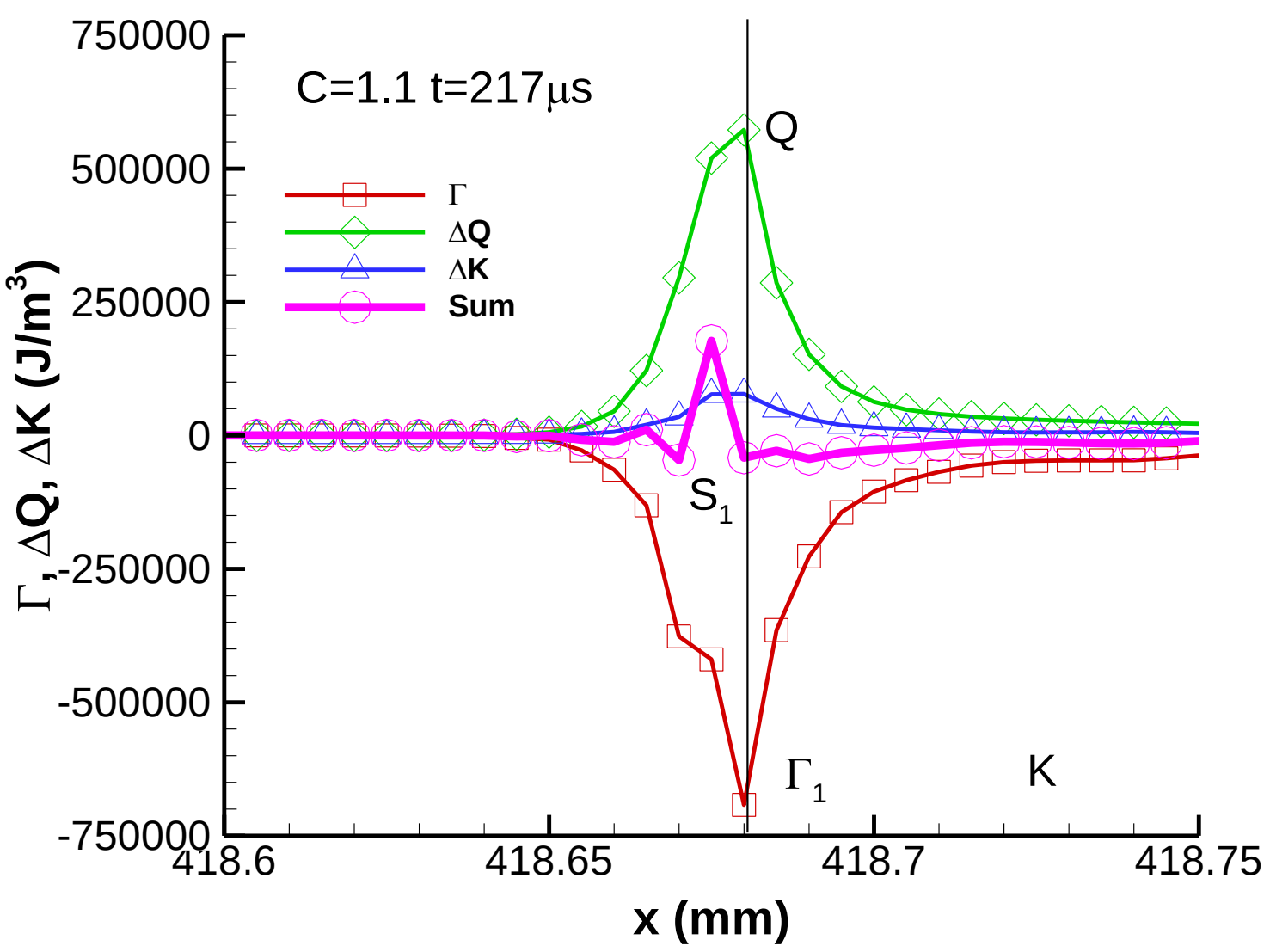

(b) the local enlarged profiles of flux

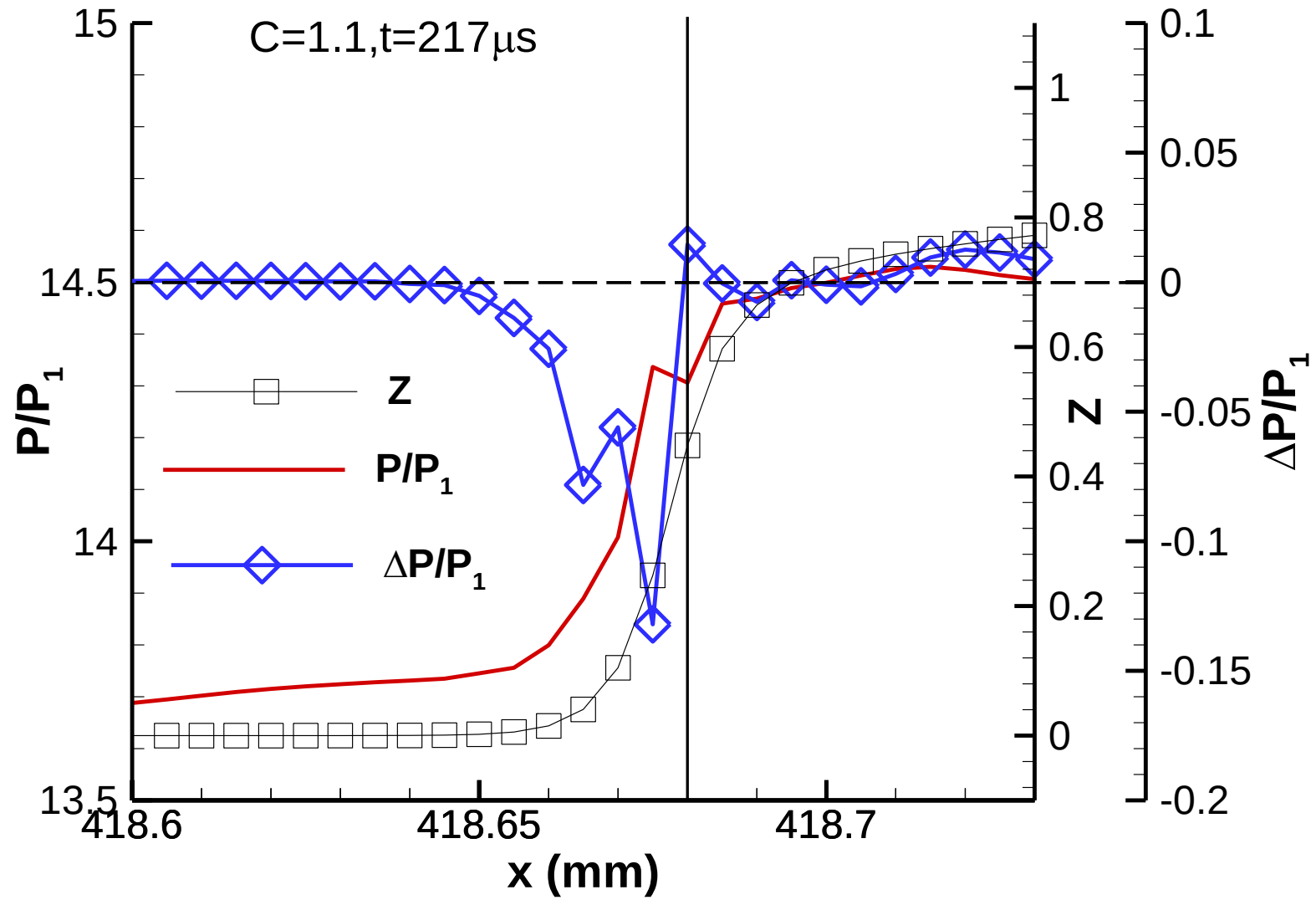

(c) the local enlarged pressure profiles

Fig.11 The profiles of flux at t=217μs of pulsating detonation

The profiles of flux at t=218μs are plotted in Fig.12. The flame is still decoupled from the shock wave at this instant. However, a new positive convective flux or a hotspot emerges in front of the flame at x=420.6mm, which is shown clearly in Fig.12(b). This local hotspot increases the total energy per unit volume due to compression waves, rather than combustion heat release. No heat is released at this hotspot at this instant because the Z profile is flat there. As a result, the local pressure and temperature are increased by this compression wave, which is shown in Fig.12(c). We can find from Fig.12(c) that the compression wave generates a positive pressure gain of about 0.16atm at x=420.6mm. Autoignition induced by this hotspot occurs at the next time step.

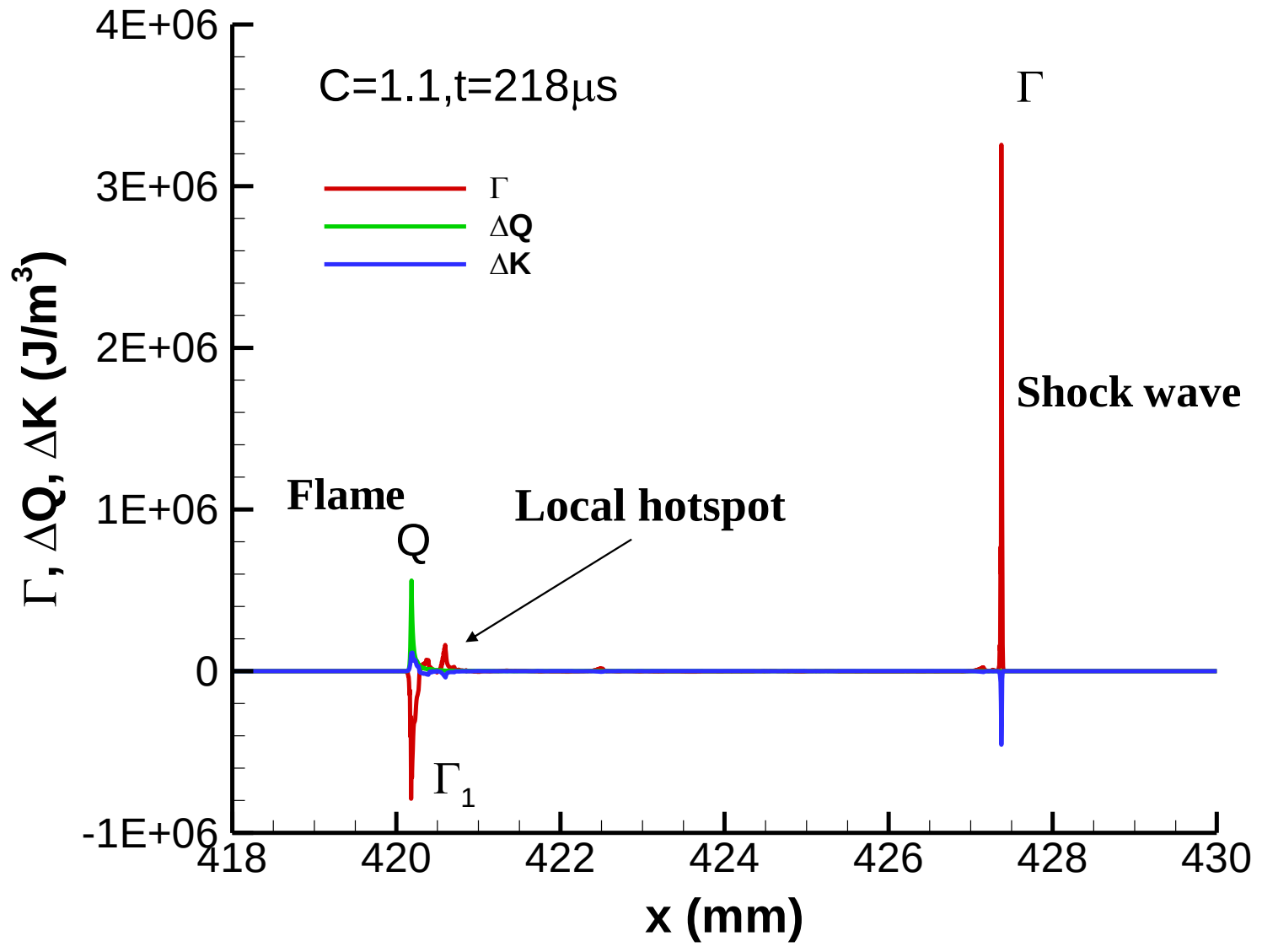

(a) the profiles of flux

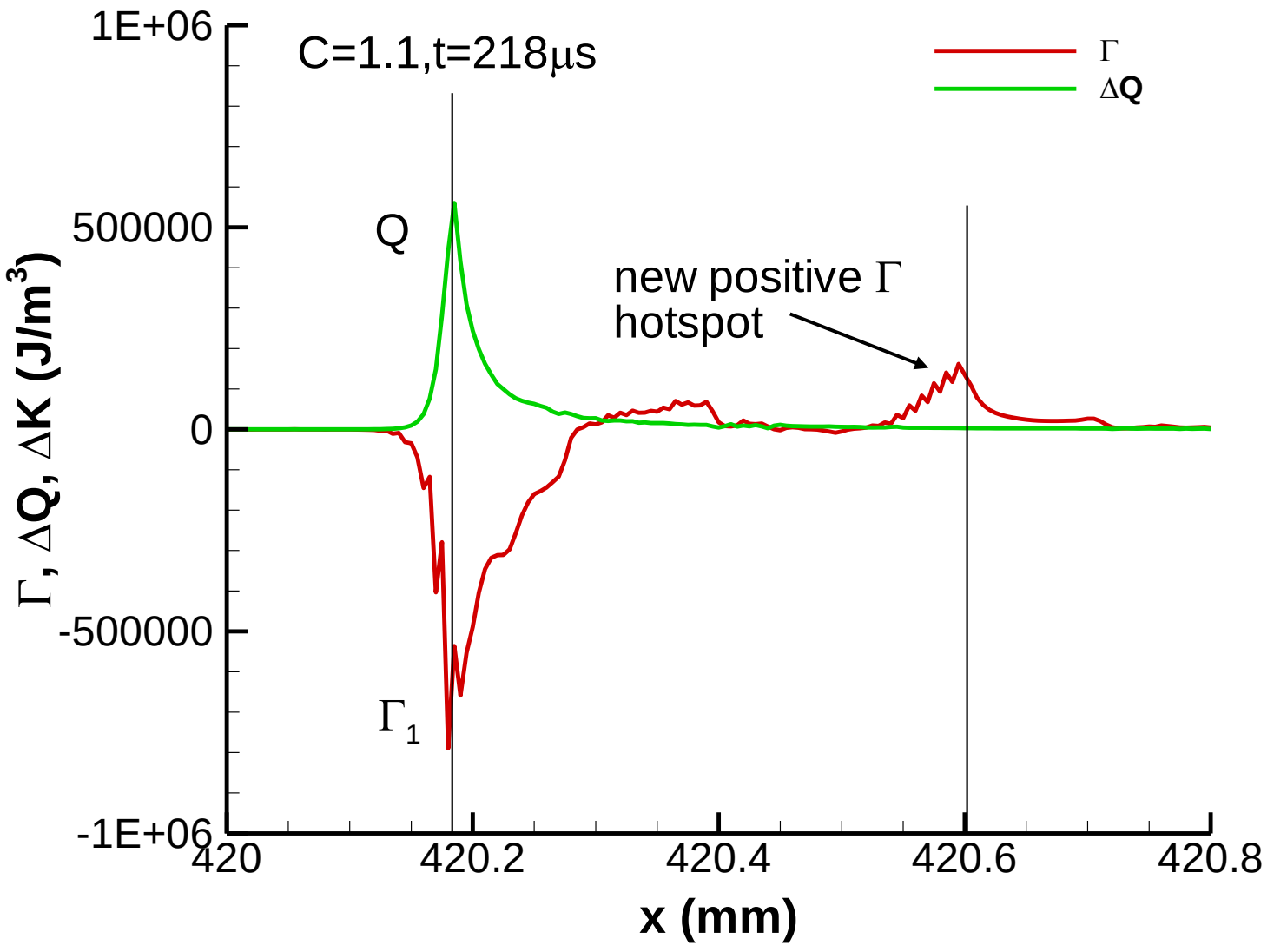

(b) the local enlarged profiles of flux

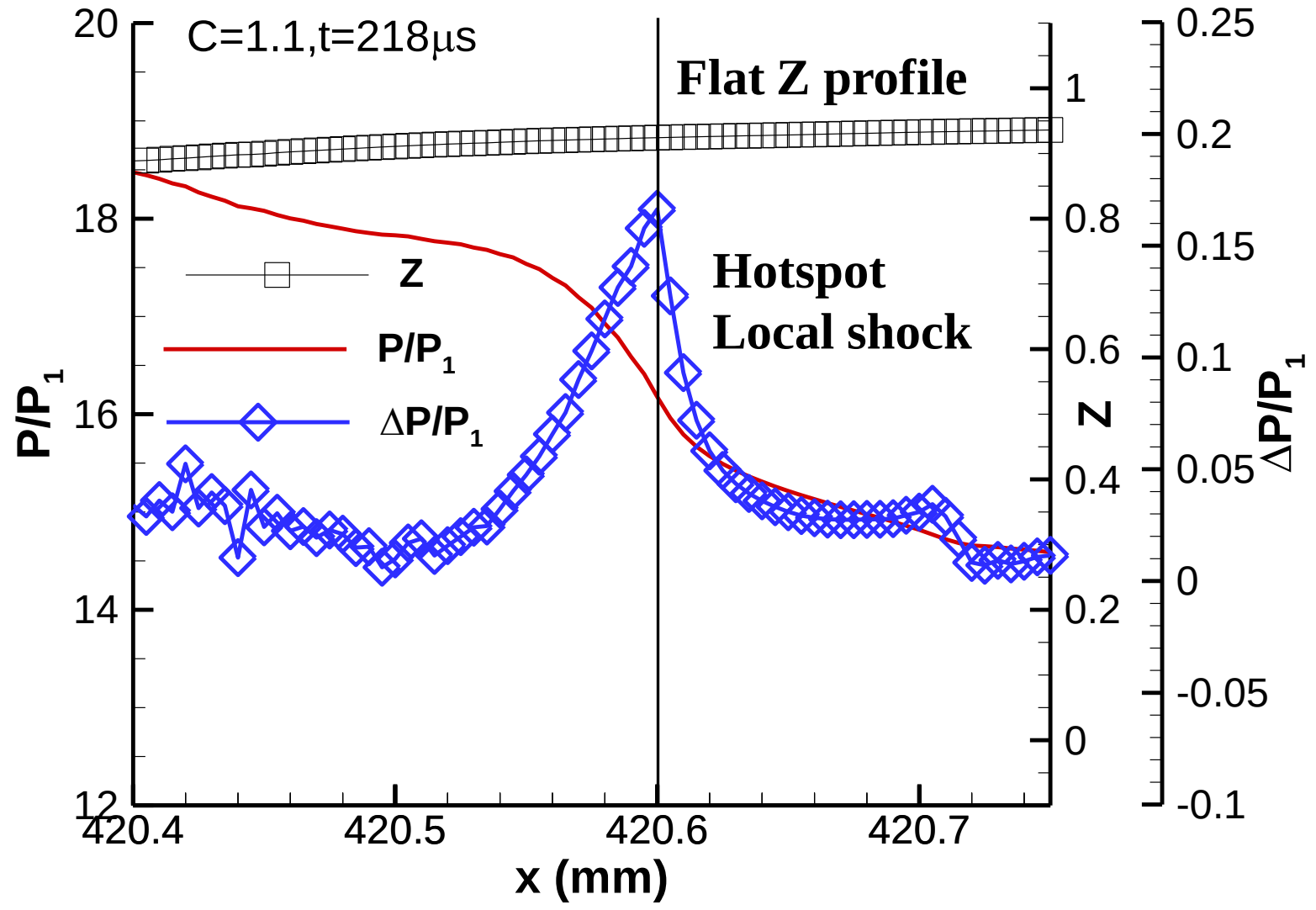

(c) local pressure increase by local hotspot

Fig.12 The profiles of flux at t=218μs of pulsating detonation

The profiles of enthalpy flux of pulsating detonation at the instant of t=217μs and 218μs are given in Fig.13. We can find from Fig.13 that the enthalpy profiles of pulsating detonation become jagged and the local slope is negative. The local negative slope of enthalpy flux at t=218μs produces a positive convective flux at about x=420.6mm, which serves as a local compression wave or a local hotspot for detonation reinitiation. This hotspot triggers detonation reignition at t=219μs, occurring at x=422.3mm, approximately 7mm behind the leading shock wave, as is shown in Fig.14. It can be seen from Fig.14(c) that the reignition produces a pressure gain of 30atm and the pressure spike increases to 120atm. An overdriven detonation is formed when this reignition detonation catches up with the leading shock wave. This is the governing mechanism of pulsating detonation.

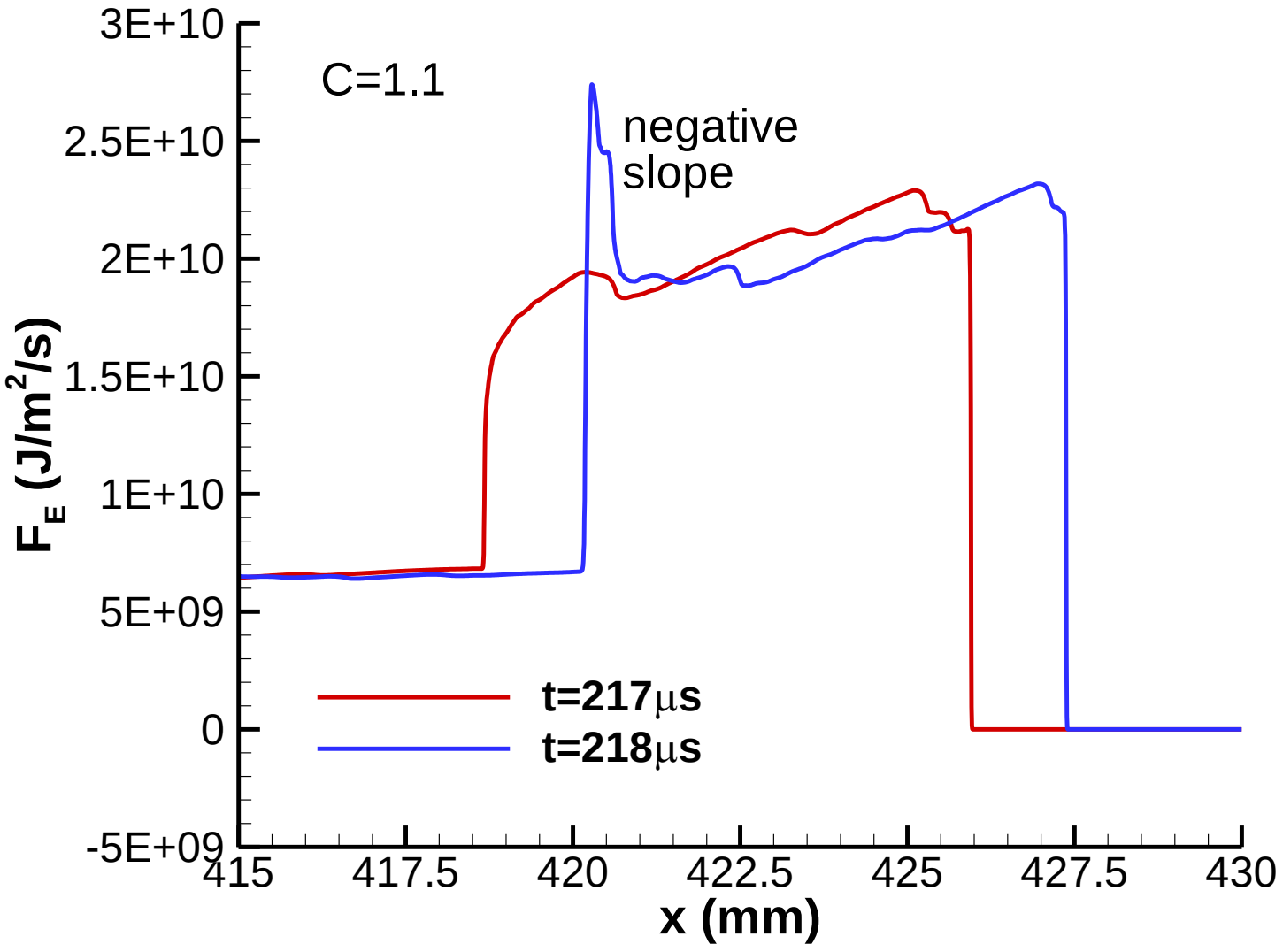

Fig.13 The profiles of enthalpy flux of pulsating detonation

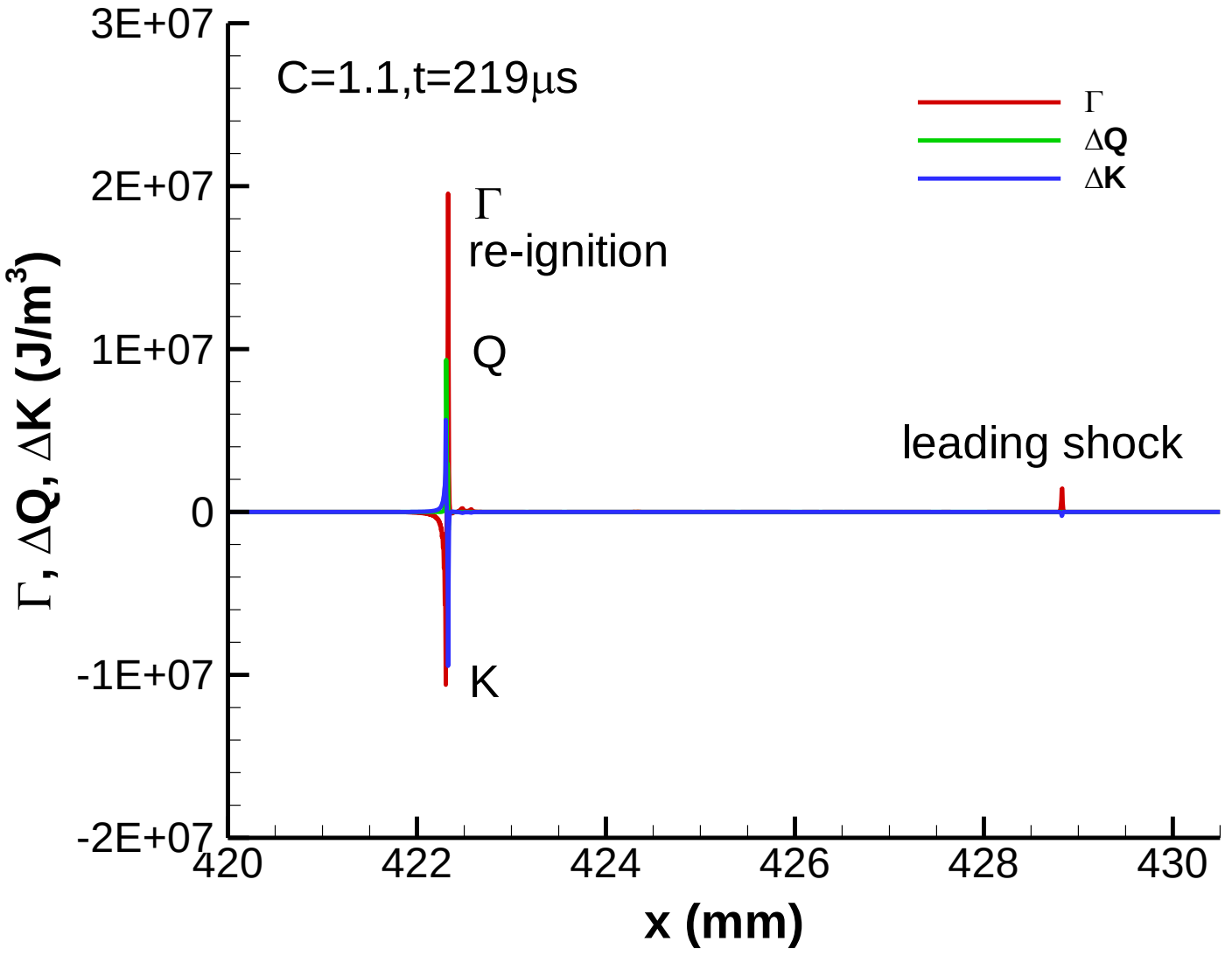

(a) the profile of flux

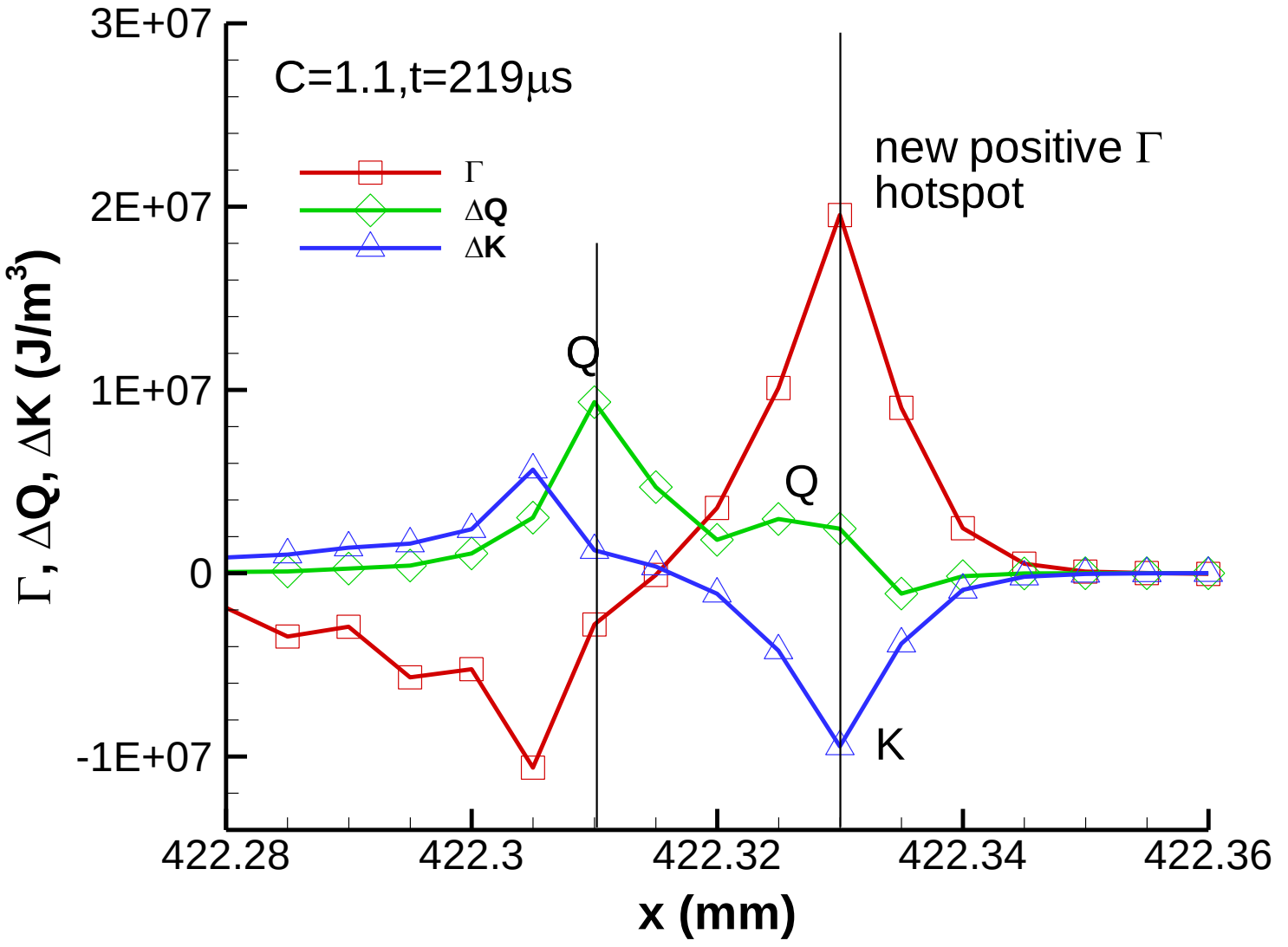

(b) local reignition by hotspot

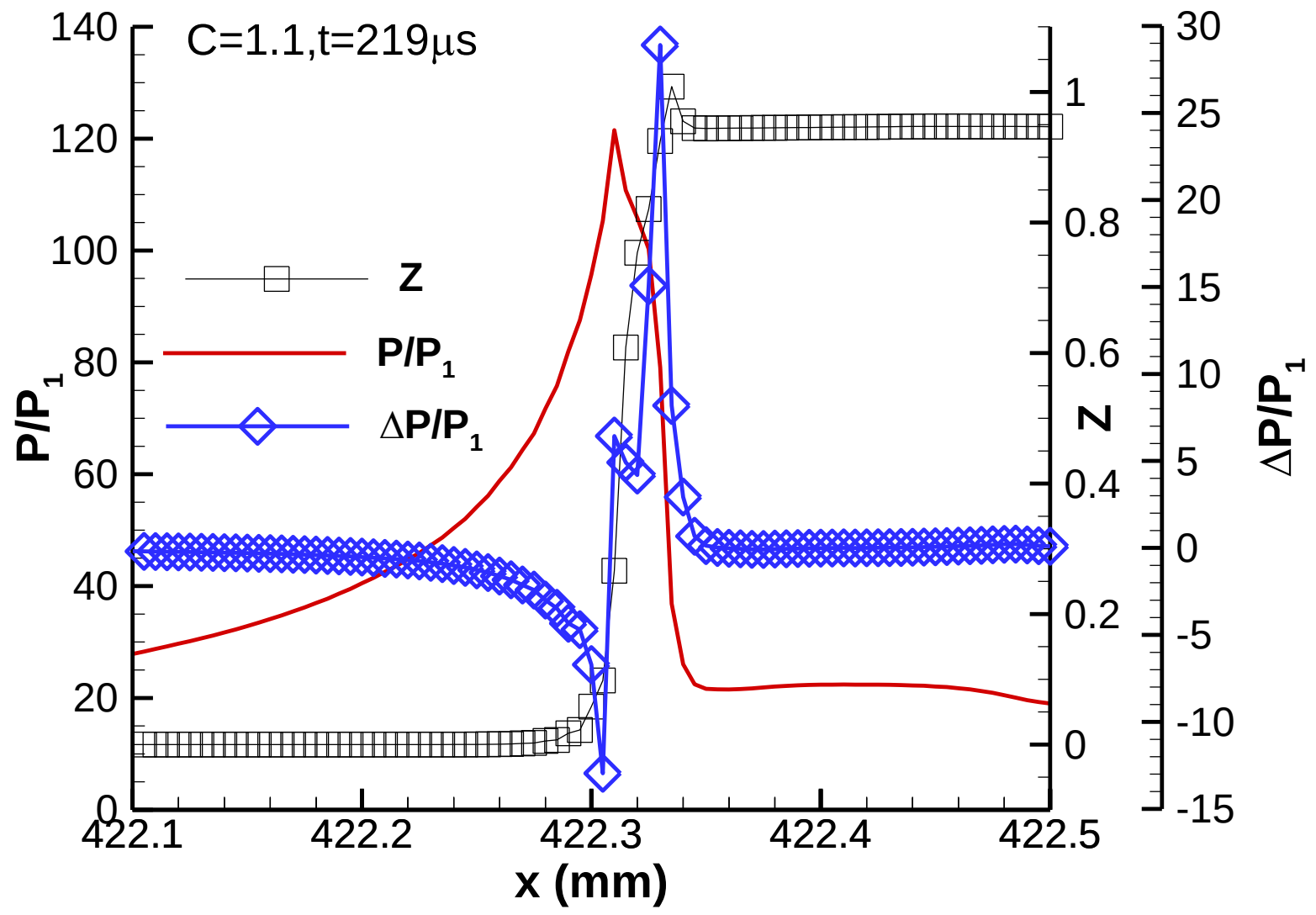

(c) local pressure profiles of autoignition

Fig.14 The profiles of flux at t=219μs of pulsating detonation

The key distinction between periodic detonation (C=1.0) and pulsating detonation (C=1.1) lies in their governing mechanisms. Periodic detonation is driven by the decoupling and recoupling of flame and shock waves, whereas pulsating detonation results from detonation quenching followed by the reignition of a new detonation wave behind a decaying leading shock wave.

### 3.6 the decaying mechanism of overdriven detonation

The formation and decaying process of overdriven detonation from t=222μs to 225μs is plotted in Fig.15. It can be seen that the reignition detonation is about 1.8mm behind the leading shock wave at t=222μs. This reignition detonation catches up with the leading shock wave and an overdriven detonation is formed at t=223μs and 224μs. This overdriven detonation reaches its highest velocity value of 2807.8m/s at this instant. However, the premature combustion occurs on the windward side of shock wave and decreases the strength of overdriven detonation gradually. Finally, the pulsating detonation becomes non-overdriven at t=225μs. The premature combustion is the decaying mechanism of an overdriven detonation.

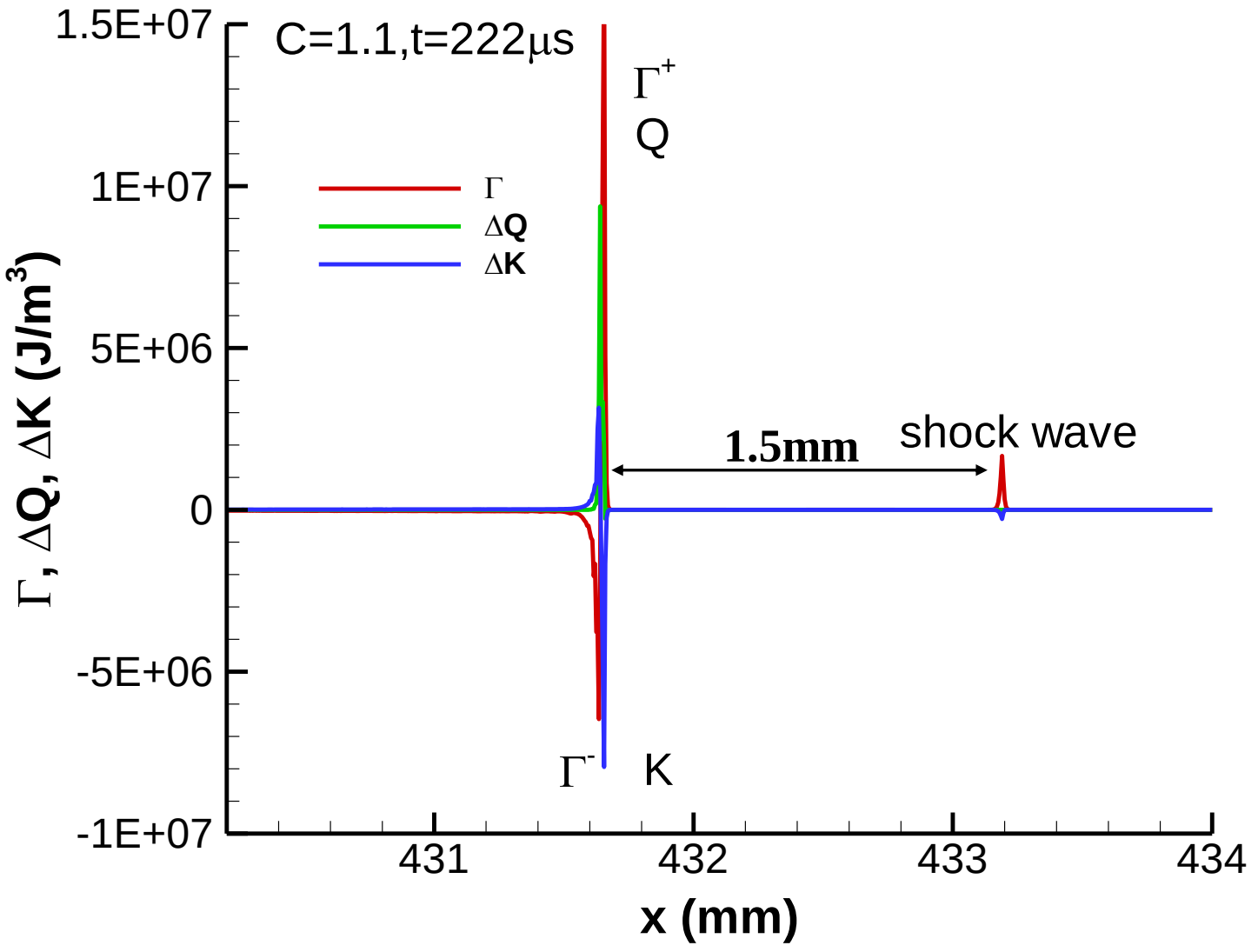

(a) t=222μs

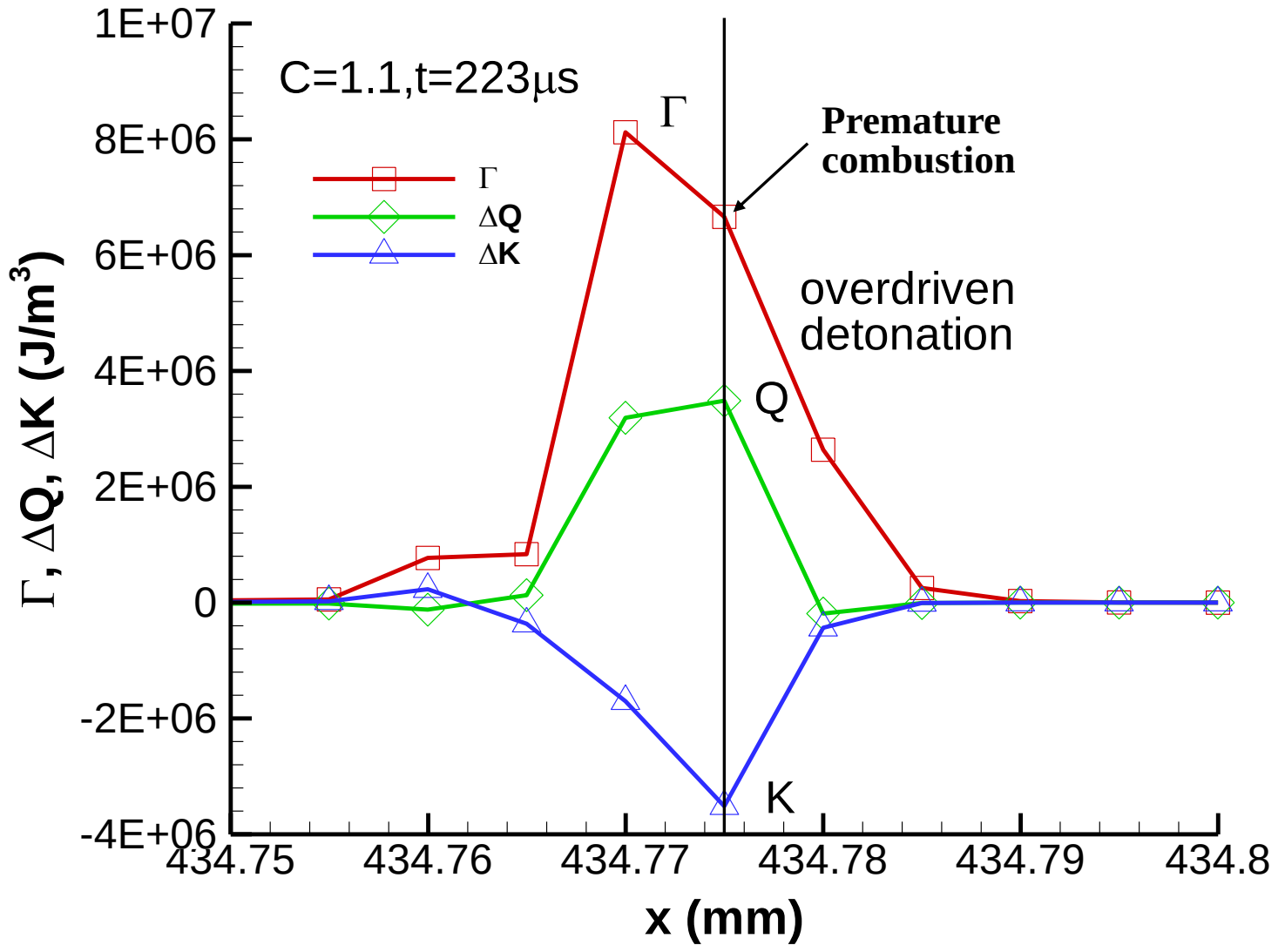

(b) t=223μs

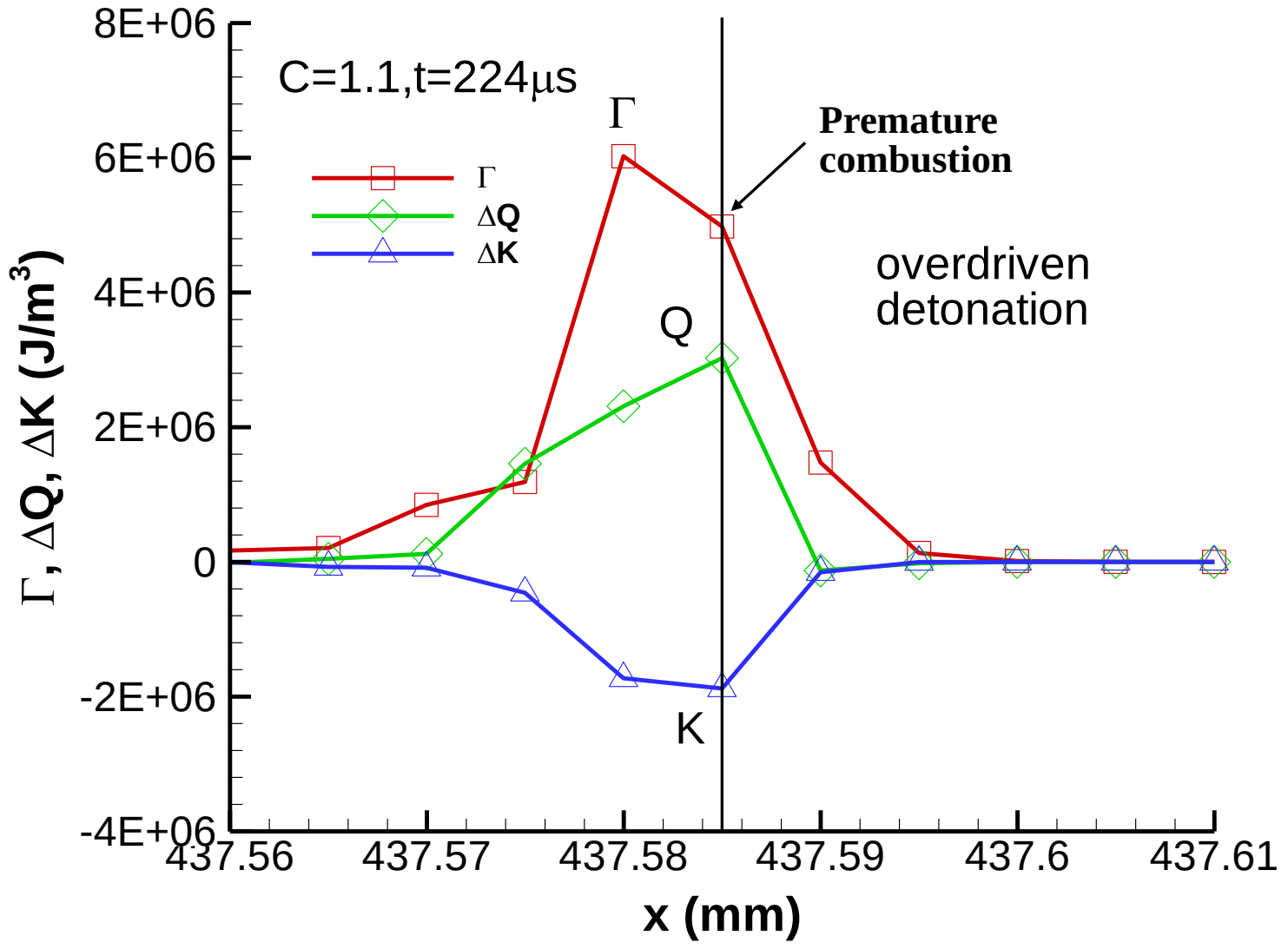

(c) t=224μs

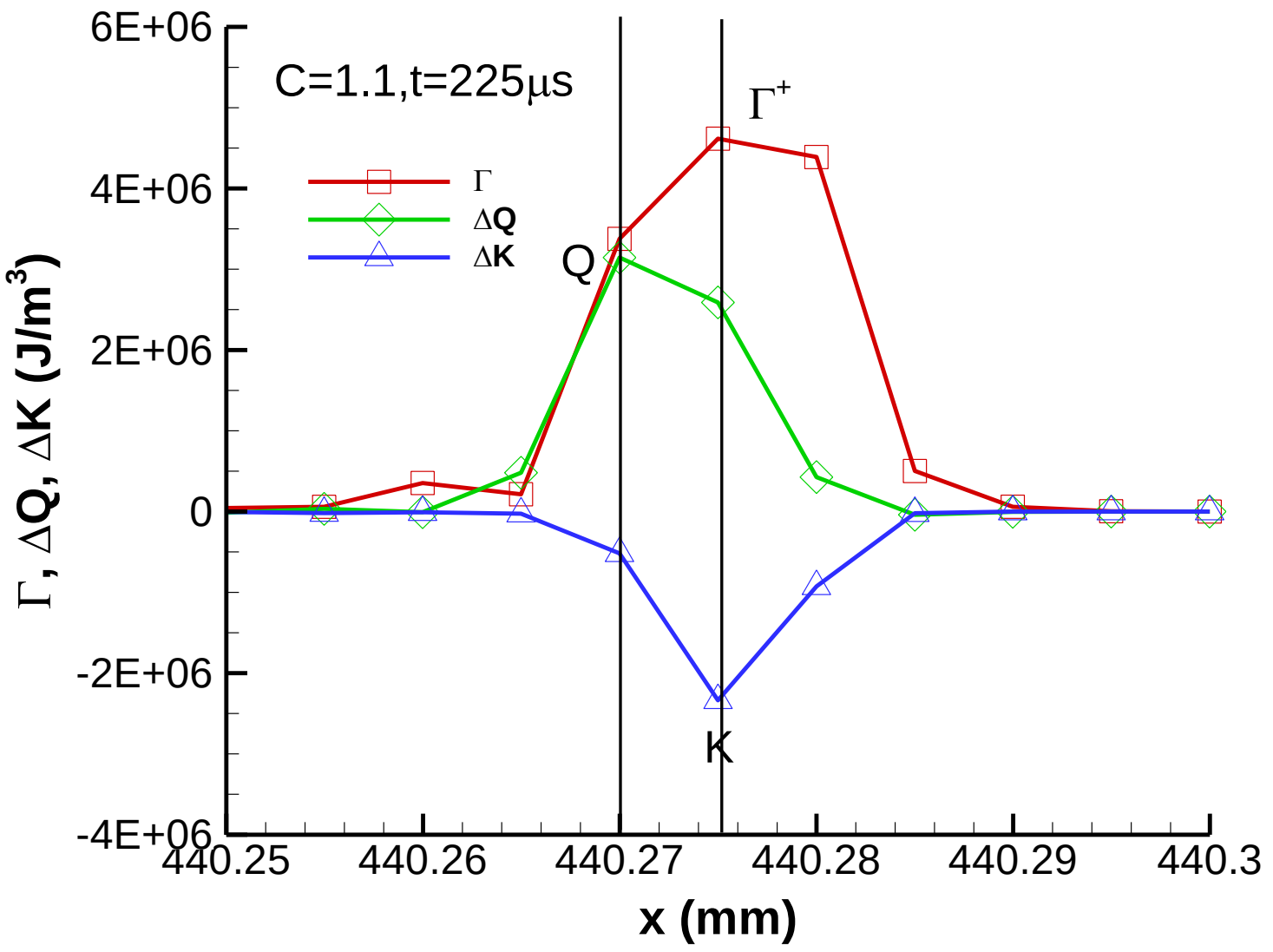

(d) t=225μs

Fig.15 The decaying process of overdriven detonation

## 4. Verification and resolution study

### 4.1 Code verification

This homemade code was used to simulate detonation successfully in previous study. In this verification study, its capacity to simulate detonation instability is checked and compared with references. Numerical simulations of Euler equations with one-step Arrhenius kinetics were conducted by Ng et al. to understand the nonlinear dynamical behavior of one-dimensional detonation [19]. Their study showed a strong similarity between one-dimensional periodic detonation and simple nonlinear dynamical systems. The maximum pressure ratio of periodic detonation wave (C=1.0) of this study (red line) are compared with that of Ref. [19] in Fig.16. In addition, the maximum pressure ratio of pulsating detonation (C=1.1) is compared with that of Ref. [26] in Fig.17. Good agreements are achieved qualitatively.

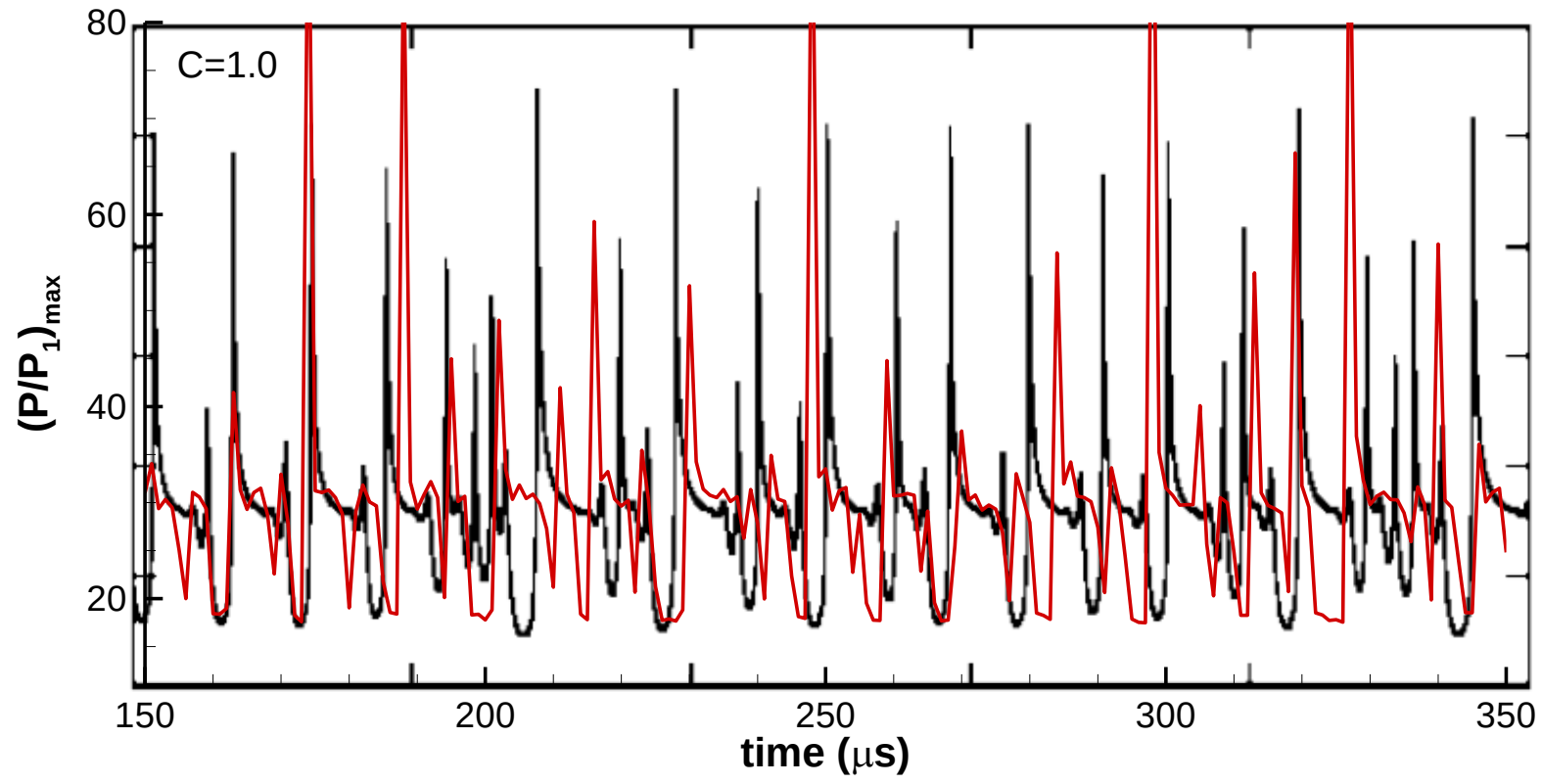

Fig.16 Comparison of maximum pressure ratio of present study (red line) with that of Ref. [19]

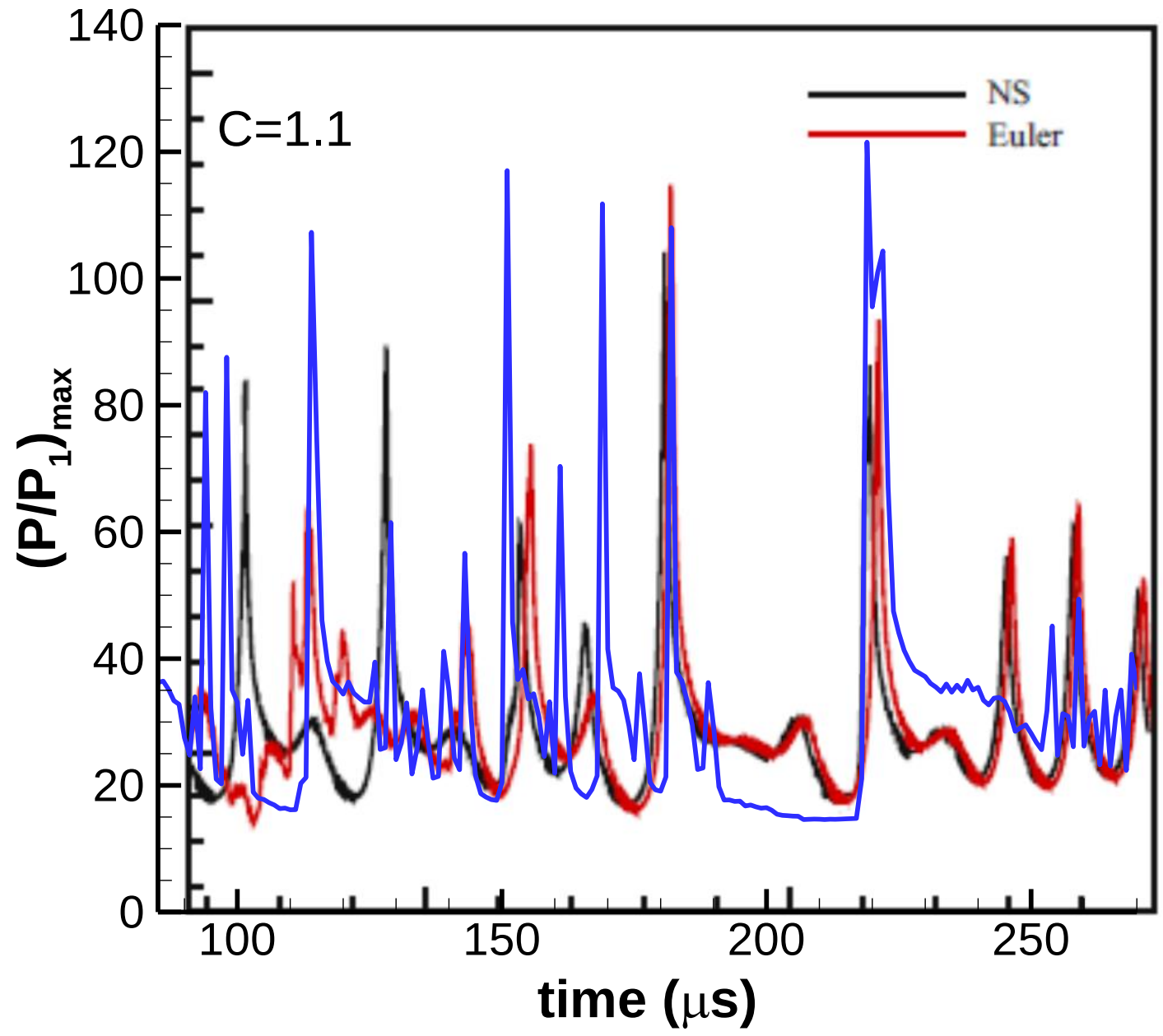

Fig.17 Comparison of maximum pressure ratio of present study (blue line) with that of Ref. [26]

### 4.2 Resolution study

This study reveals that the convective flux plays an important role in the detonation instability. The aim of this resolution study is to demonstrate that this conclusion holds under different grid sizes. The resolution study with uniform gird size of 2.5μm was conducted and the velocity under different activation energy are plotted in Fig.18. It can be seen that the three regions of detonation propagation mode: a stable region (C=0.8–0.9), an unstable region (C=1.0–1.1), and a quenching region (C≥1.15), are reproduced successfully.

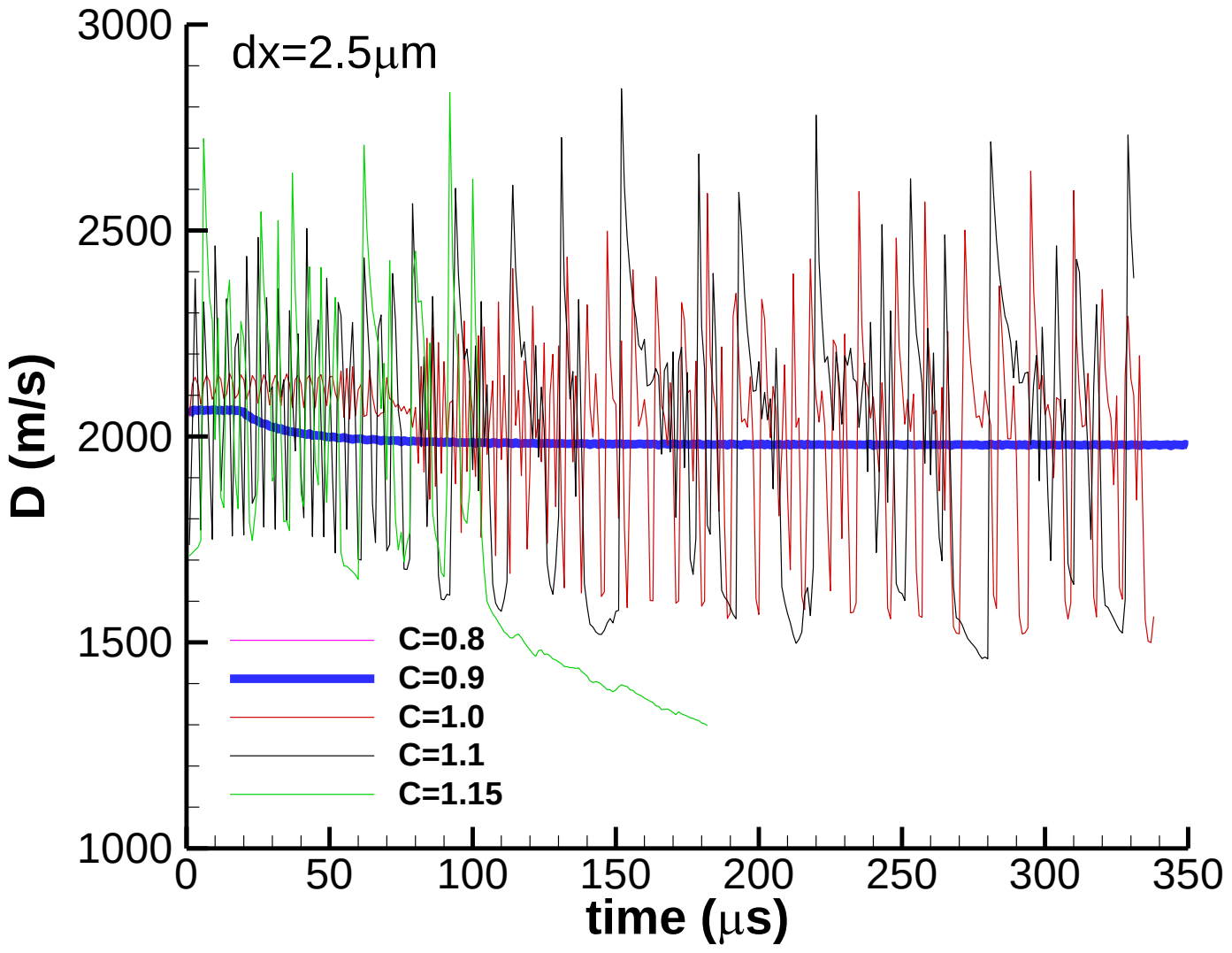

(a) detonation velocity under different activation energy

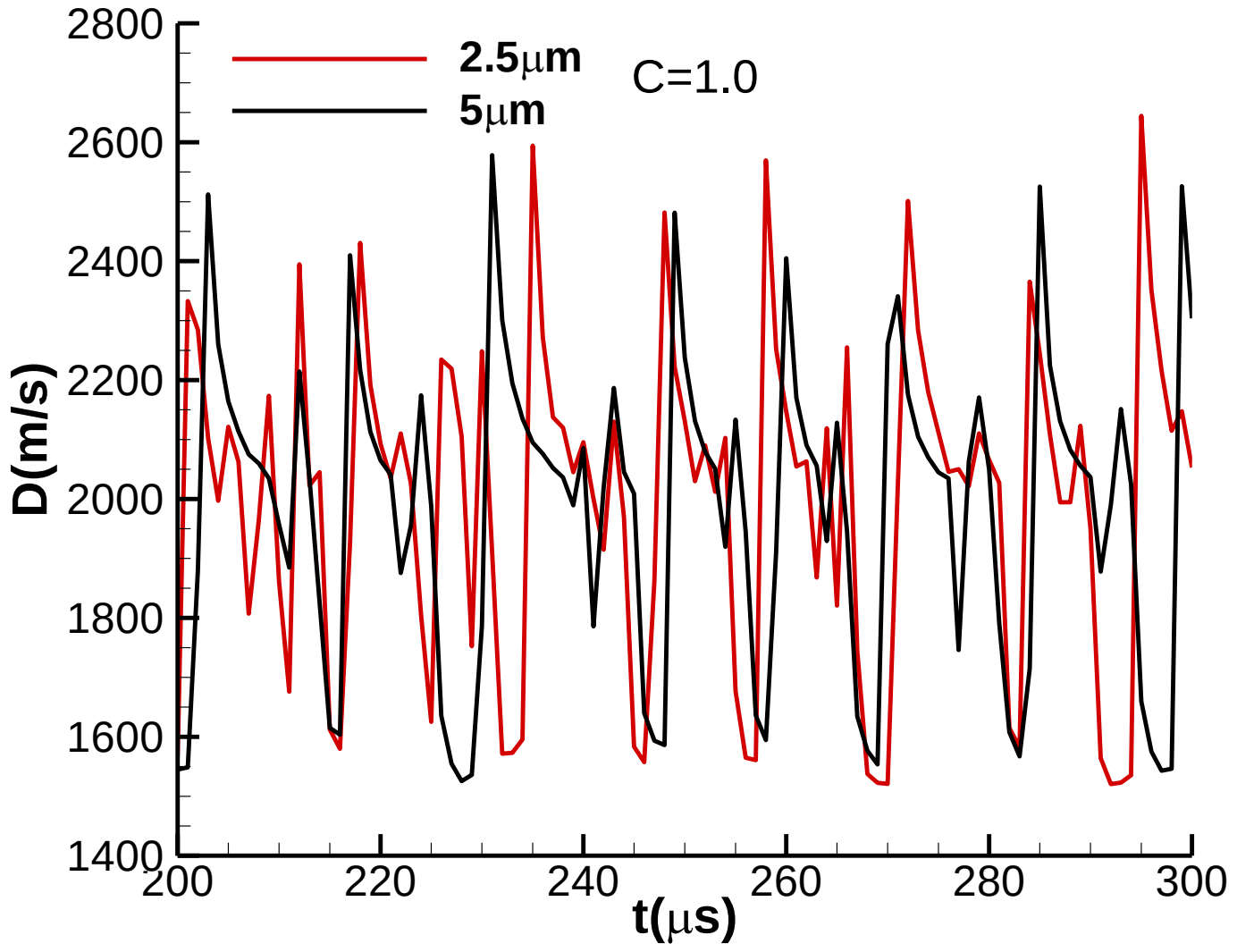

(b) velocity comparison with different grid size

Fig.18 Detonation velocity with grid size of 2.5μm

## 5. Conclusions

One-dimensional numerical simulations with Euler equations and irreversible one-step Arrhenius kinetics are performed to study the underlying physics of the instability of one-dimensional gaseous detonation. The theoretical equation of pressure gain of detonation products within one time step is deduced. The stable detonation, the pulsating detonation, and the detonation quenching process are obtained successfully by increasing the activation energy in the numerical simulations. This study comes to the following conclusions.

(1) The theoretical pressure gain equation identifies three key parameters governing detonation propagation: convective flux, kinetic energy flux, and chemical reaction heat release. The pressure gain or the internal energy gain of detonation products are attributed to the algebra summation of these three fluxes.

(2) The detonation is stable under lower activation energy. The heat release flux is coupled with the shock wave tightly and almost 100% heat is released on the windward side of shock wave because of the very fast chemical reaction rate under lower activation energy. The stable detonation is constant volume combustion.

(3) The detonation becomes unstable, pulsating and chaotic with the increase of activation energy. The flame is completely decoupled from the shock wave for the unstable detonation. The negative convective flux at the flame front reduces the total energy per unit volume of combustion products, effectively resulting in constant pressure combustion mode.

(4) The negative convective flux plays a key role in the detonation instability and

detonation quenching process, which decreases the internal energy of detonation products. The negative convective flux is produced by the sharper and steeper enthalpy flux at the detonation front, whose slope becomes very big once the flame is decoupled from the shock wave.

(5) The shaper and steeper enthalpy flux is produced by the significant variation of physical parameters on the von Neumann and on the C-J detonation point. Once the flame decouples from the shock wave, the enthalpy flux on the von Neumann spike increases by an order of magnitude, inducing a very stronger negative convective flux or rarefaction wave, which suppresses the pressure gain generated by heat release.

11900-11908.